\documentclass[3p,times,twocolumn]{elsarticle}

\usepackage{ecrc}

\volume{00}

\firstpage{1}

\journalname{Nuclear Physics B Proceedings Supplement}

\runauth{Roman Kogler}

\jid{nuphbp}

\jnltitlelogo{Nuclear Physics B Proceedings Supplement}

\usepackage{amssymb}
\usepackage{amsmath}
\usepackage{bm}
\usepackage{color}

\usepackage[figuresright]{rotating}

\usepackage{xspace}
\usepackage{acronym}

\newcommand{\as}{\ensuremath{\alpha_s}\xspace}
\newcommand{\asmz}{\ensuremath{\alpha_s(M_Z)}\xspace}
\newcommand{\asmu}{\ensuremath{\alpha_s(\mu_{\mathrm{r}})}\xspace}

\newcommand{\xbj}{\ensuremath{x}\xspace}
\newcommand{\Qsq}{\ensuremath{Q^2}\xspace}

\newcommand{\muf}{\ensuremath{\mu_{\mathrm{f}}}\xspace}
\newcommand{\mur}{\ensuremath{\mu_{\mathrm{r}}}\xspace}

\newcommand{\MeanPt}{\ensuremath{\langle P_{\mathrm{T}} \rangle}\xspace}
\newcommand{\Pt}{\ensuremath{P_{\mathrm{T}}}\xspace}
\newcommand{\Ptone}{\ensuremath{P_{\mathrm{T,1}}}\xspace}
\newcommand{\Pttwo}{\ensuremath{P_{\mathrm{T,2}}}\xspace}
\newcommand{\kt}{\ensuremath{k_{\mathrm{T}}}\xspace}

\newcommand{\etalab}{\ensuremath{\eta_{\mathrm{lab}}}\xspace}
\newcommand{\Mjj}{\ensuremath{M_{\mathrm{12}}}\xspace}

\newcommand{\xgamma}{\ensuremath{x_\gamma}\xspace}
\newcommand{\xgammaobs}{\ensuremath{x_\gamma^{\mathrm{obs}}}\xspace}

\newcommand{\ud}{\ensuremath{\mathrm{d}}\xspace}

\newcommand{\Oa}{\ensuremath{\mathcal{O}(\alpha_s)}\xspace}
\newcommand{\Oaa}{\ensuremath{\mathcal{O}(\alpha_s^2)}\xspace}
\newcommand{\Oaaa}{\ensuremath{\mathcal{O}(\alpha_s^3)}\xspace}

\newcommand{\Pth}{\ensuremath{P_{\mathrm{T}}^{\mathrm{h}}}\xspace}
\newcommand{\Ptda}{\ensuremath{P_{\mathrm{T}}^{\mathrm{da}}}\xspace}

\newcommand{\eq}{Eq}
\newcommand{\fig}{Fig}

\begin{document}


\acrodef{BBE}{Backward Barrel\acroextra{, Electromagnetic (calorimeter wheel)}}
\acrodef{BCDMS}{Bologna-Cern-Dubna-Munich-Saclay}
\acrodef{BGF}{boson gluon fusion}
\acrodef{BPC}{Backward Proportional Chamber}
\acrodef{BST}{Backward Silicon Tracker}
\acrodef{CC}{charged current}
\acrodef{CDM}{colour dipole model}
\acrodef{CB}{Central Barrel\acroextra{ (calorimter wheel)}}
\acrodef{CIP}{Central Inner Proportional Chamber}
\acrodef{CJC}{Central Jet Chamber}
\acrodef{COP}{Central Outer Proportional Chamber}
\acrodef{COZ}{Central Outer z-Chamber}
\acrodef{CST}{Central Silicon Tracker}
\acrodef{CTD}{Central Track Detector}
\acrodef{DESY}{Deutsches Elektronen Synchrotron}
\acrodef{DIS}{deep-inelastic scattering}
\acrodef{DGLAP}{Dokshitzer, Gribov, Lipatov, Altarelli, Parisi}
\acrodef{DREAM}{Dual-Readout Module}
\acrodef{DST}{Data Summary Tape}
\acrodef{DVCS}{deeply virtual compton scattering}
\acrodef{EW}{electroweak}
\acrodef{EMC}{electromagnetic calorimeter}
\acrodef{ET}{Electron Tagger}
\acrodef{FB}{Forward Barrel\acroextra{ (calorimeter wheel)}}
\acrodef{FMD}{Forward Muon Detector}
\acrodef{FST}{Forward Silicon Tracker}
\acrodef{FTD}{Forward Track Detector}
\acrodef{FTT}{Fast Track Trigger}
\acrodef{H1OO}{H1 Object Oriented Analysis Software}
\acrodef{HAC}{hadronic calorimeter}
\acrodef{HAT}{H1 Event Tag}
\acrodef{HERA}{Hadron-Elektron-Ring-Anlage}
\acrodef{HFS}{hadronic final state}
\acrodef{IF}{Inner Forward\acroextra{ (calorimeter wheel)}}
\acrodef{IP}{interaction point}
\acrodef{IR}{infrared}
\acrodef{LAr}{liquid argon}
\acrodef{LEP}{Large Electron Positron Collider}
\acrodef{LHC}{Large Hadron Collider}
\acrodef{LO}{leading order}
\acrodef{MC}{Monte Carlo\acroextra{ (event generator)}}
\acrodef{MEPS}{matrix elements and parton shower}
\acrodef{NC}{neutral current}
\acrodef{NMC}{New Muon Collaboration}
\acrodef{NNLO}{next-to-next-to-leading order}
\acrodef{NLO}{next-to-leading order}
\acrodef{OF}{outer forward\acroextra{ (calorimeter wheel)}}
\acrodef{PETRA}{Positron-Elektron-Ring-Anlage}
\acrodef{PD}{photon detector}
\acrodef{PDF}{parton distribution function}
\acrodef{POT}{Production Output Tape}
\acrodef{pQCD}{perturbative \ac{QCD}}
\acrodef{QCD}{quantum chromodynamics}
\acrodef{QCDC}{\ac{QCD} Compton}
\acrodef{QED}{quantum electrodynamics}
\acrodef{QEDC}{\ac{QED} Compton}
\acrodef{QPM}{quark parton model}
\acrodef{RGE}{renormalisation group equation}
\acrodef{SLAC}{Stanford Linear Accelerator Center}
\acrodef{SM}{Standard Model}
\acrodef{SpaCal}{Spaghetti Calorimeter}
\acrodef{TC}{Tail Catcher}
\acrodef{ToF}{Time-of-Flight}
\acrodef{UV}{ultraviolet}


\begin{frontmatter}

\dochead{} 

\title{Precision jet measurements at HERA and determination of $\alpha_s$\tnoteref{t1}}
\tnotetext[t1]{Presented at the Ringberg Workshop \emph{New Trends in HERA Physics}, September 25--28, 2011 (to be published in Nucl.\ Phys.\ B Proc.\ Suppl.).}

\author{Roman Kogler}
\ead{roman.kogler@desy.de}
\author{for the H1 and ZEUS Collaborations}
\address{Deutsches Elektronen-Synchrotron (DESY), Notkestra\ss{}e 85, 22607 Hamburg, Germany}

\begin{abstract}
\acused{QCD}

The status is reviewed of recent high precision measurements of inclusive-jet, dijet and trijet production in deep-inelastic scattering and photoproduction by the HERA experiments H1 and ZEUS. The measurements are in good agreement with \acl{pQCD} calculations at \acl{NLO} and are used for the extraction of the value of the strong coupling at the mass of the $Z$ boson, \asmz.  The methods and results of the \ac{QCD} analyses are presented and a summary of the values of \asmz from recent jet measurements at HERA is given. 

\end{abstract}

\begin{keyword}
deep-inelastic scattering \sep jet production \sep quantum chromodynamics  \sep strong coupling
\PACS 13.60.Hb \sep 13.87.Ce \sep 12.38.Qk
\end{keyword}

\end{frontmatter}

\acresetall 

\section{Introduction}
\label{sec:intro}

Since the advent of particle colliders the study of jet production has been essential for our understanding of the strong force and the development of \ac{QCD}. Nowadays \ac{pQCD} is firmly established and jets are used extensively at the Tevatron and the \ac{LHC} in analyses searching for new physical phenomena. In these analyses an overwhelming amount of background originates from \ac{SM} \ac{QCD} processes. Therefore potential discoveries depend heavily on our understanding of \ac{QCD}, which can be improved by precision jet measurements. 

One of the key predictions of \ac{QCD} is the running of the strong coupling \as as function of the renormalisation scale \mur. However, the absolute value of \as at some starting scale has to be obtained from experimental data. The direct sensitivity of the jet production cross sections to \as makes jets an ideal tool for precision measurements of \as. Additionally, since jet data usually span over a large range of values of \mur, it can be used for rigourous tests of the evolution of \as. 

In contrast to hadron-hadron collisions, in \ac{DIS} there is only one proton present in the initial state. Therefore \ac{DIS} provides a very clean environment for jet production. It is possible to study \ac{pQCD} without the complication of having to disentangle contributions from pile-up and multiple interactions, and only a minimum amount of modelling of non-perturbative effects is needed.

Inclusive \ac{NC} and \ac{CC} \ac{DIS} data are used in analyses of the proton structure for determinations of the \acp{PDF} of the proton. These are one of the key ingredients for predictions at the LHC and uncertainties on the \acp{PDF} are reflected in the theoretical predictions. Inclusive \ac{DIS} data are sensitive to the gluon density and \as only in \ac{NLO} through scaling violations, however. Jet data provide direct sensitivity to the gluon density and \as already in \ac{LO}. The inclusion of \ac{DIS} jet data reduces the correlation between the gluon \ac{PDF} and \as, thereby making a simultaneous determination of proton \acp{PDF} and \as with \ac{DIS} data alone possible~\cite{H1prelim-11--034}.

Jet production in photoproduction, where a proton collides with a quasi real photon, additionally provides information on the structure of the photon. Since in these interactions the photon can show hadronic structure, this kinematic regime closes the gap between \ac{DIS} and hadron-hadron collisions. Having obtained the proton \acp{PDF} using \ac{DIS} data, jets in photoproduction can also be used for independent tests of the obtained proton \acp{PDF}. Additionally, in photoproduction the spectator partons in the proton not participating in the primary interaction, may interact with partons of the resolved photon. In these processes multiple interaction models can be studied in an environment free of pile-up. 

The HERA collider provides a unique opportunity to study jet production in electron-proton ($ep$) collisions at the highest available centre-of-mass energies. After its shutdown in the year 2007 the H1 and ZEUS collaborations invested a lot of effort in improving even further the detector calibrations and reconstruction algorithms. The benefits of these efforts become visible now with jet measurements of unprecedented precision. In this document the most recent precision jet measurements at HERA are reviewed, together with the \ac{QCD} analyses and determinations of \as from jet cross sections. 
 
\section{QCD and the running coupling}

One of the central predictions of \ac{QCD} is the dependence of the strong coupling on the renormalisation scale \mur, which can be obtained from the renormalisation-group equations. It is expressed in terms of the $\beta$-function
\begin{equation}
\beta(\asmu) = \mur \frac{\partial \asmu}{\partial \mur} \, , 
\label{eq:alphas_running}
\end{equation}
which can be expanded in powers of \as and the resulting coefficients $\beta_n$ are known up to four loops~\cite{Ritbergen:97:379}. The behaviour of \as as function of the scale \mur is one of the best-known quantities in \ac{QCD}. However, the absolute value of \as at some starting scale $\mu_0$ has to be obtained from experiment. Usually the choice $\mu_0 = M_Z$ is used, where $M_Z$ is the mass of the $Z$-boson. Uncertainties in the value of \asmz are reflected in uncertainties of \ac{SM} predictions at the \ac{LHC} such as $W$, $Z$, $t \bar{t}$ or Higgs production. The spread of the predicted cross sections, usually combined with the uncertainty from proton \ac{PDF} parametrisations, can be as large as 10\%~\cite{Watt:11:69}, showing the need for precision measurements of \asmz. Additionally, since one of the key predictions of \ac{QCD} is the universality of \asmz, testing the compatibility of the obtained values of \asmz from a variety of different processes measured at different scales \mur, provides a stringent test of \ac{QCD} itself.

\section{Deep-Inelastic Scattering}

\begin{figure}
\begin{center}
\includegraphics[width=7.8cm]{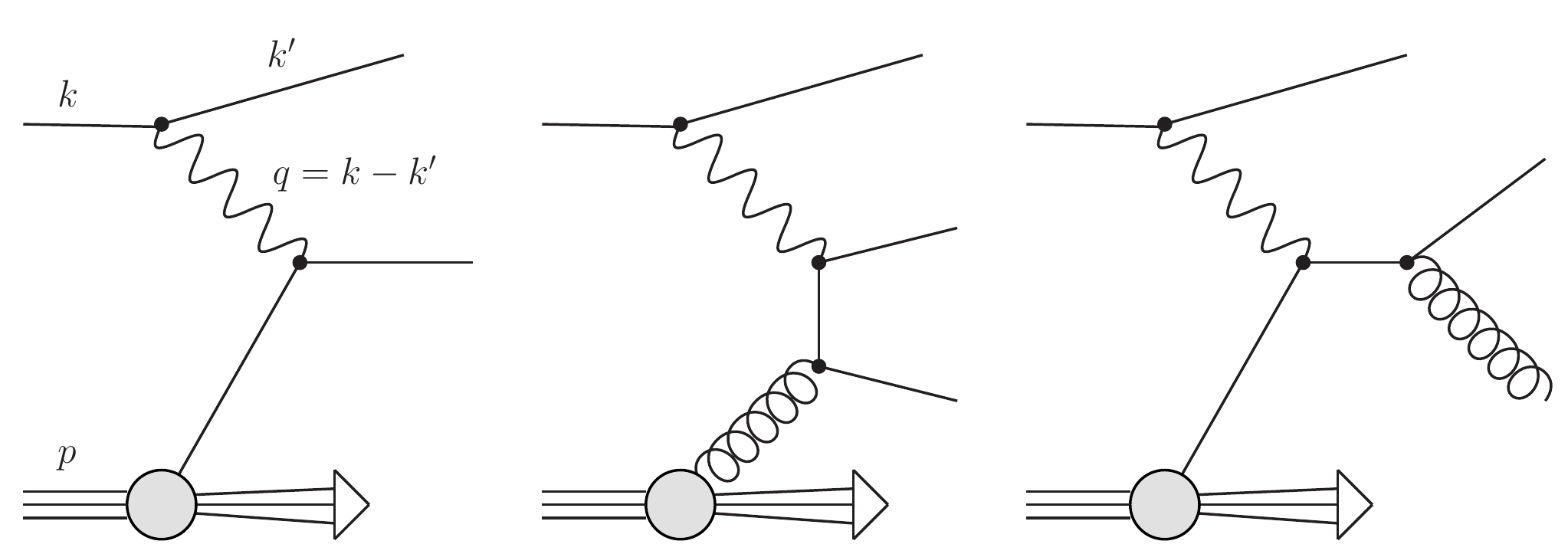}
\end{center}
\vspace{-0.3cm}
\caption{\label{fig:feyn} Some important Feynman diagrams of neutral current $ep$-scattering. The \ac{LO} process is shown on the left, followed by one in each case of the boson-gluon fusion and QCD Compton diagrams. }
\end{figure}

The \ac{LO} Feynman diagram of $ep$ scattering is shown in Figure \ref{fig:feyn} (left). The incoming electron with four-momentum $k$ interacts with the proton with four-momentum $P$ through the exchange of a virtual boson with four-momentum $q=k-k'$. In \ac{NC} reactions, where the exchanged boson is either a photon or a $Z$-boson, the final state consists of a scattered electron with four-momentum $k'$ and the hadronic final state $X$. 

The kinematics of the scattering process can be described by the negative four-momentum transfer squared $\Qsq=-q^2$, also called the virtuality of the exchanged photon, the Bjorken-scaling variable $\xbj = \Qsq / (2 P \cdot q)$ and the inelasticity $y=(P \cdot q)/(P \cdot k)$. These variables are related at fixed centre-of-mass energy $\sqrt{s}$, through the relation $\Qsq = s x y$, such that two variables are sufficient to uniquely define the scattering process. 

The variable \Qsq is also used to separate the $ep$ scattering process into two different kinematic regimes. At vanishing \Qsq the proton collides with a quasi-real photon and this $\gamma^* p$ interaction is referred to as photoproduction. For \Qsq significantly larger than 1~GeV$^2$ and the invariant mass of $X$ much larger than the proton mass, one speaks of \ac{DIS}.

The inclusive \ac{NC} scattering cross section for the reaction $e^{\pm} + p \to e^{\pm} + X$ can be calculated in terms of the structure functions $F_2$, $\xbj F_3$ and $F_L$~\cite{Aaron:10:109}. The structure function $F_2$ gives the dominating contribution to the cross section in the bulk of the phase space. It is in \ac{LO} related to the quark densities of the proton, $F_2 = x \sum_i e_i^2 (q_i + \bar{q}_i)$, where the sum runs over all active quark flavours, and the charge $e_i$ of quark $i$ is given in units of the elementary charge $e$. The structure function $\xbj F_3$ is directly proportional to the difference between the quark and anti-quark densities, and $F_L$ is sensitive to the gluon density. However, contributions to the cross section from the gluon density, either directly through $F_L$ or because of corrections to $F_2$ (scaling violations), are non-zero only at \ac{NLO} and beyond. Furthermore, at \ac{NLO} the gluon density enters the DGLAP equations multiplied by \as, which leads to a large correlation between the gluon density and \as in determinations of \acp{PDF} of the proton.

\section{Jet production at HERA}

While the inclusive $ep$ scattering process is at \ac{LO} purely of electroweak nature, and \ac{QCD} enters only through higher-order corrections, in jet production \ac{QCD} plays a dominant role already at \ac{LO}. This leads to a direct sensitivity of the jet cross sections to the strong coupling \as and the gluon density of the proton. Hence, the study of jet production provides a prominent way to determine the value of \asmz as well as important complementary information for \ac{QCD} analyses of structure function data~\cite{HERAPDF}. 

Jets are collimated sprays of hadrons, which are created by the showering and hadronisation of outgoing final state partons. The exact definition of a jet depends on the employed jet algorithm, which specifies which hadrons are clustered into a jet and how the jet four-momentum is calculated. At HERA, usually the longitudinally invariant \kt algorithm~\cite{Catani:93:187, Ellis:93:3160} with the \Pt recombination scheme \cite{Huth:90} is used. Recently also studies with the anti-\kt~\cite{Cacciari:08:063} and the SISCone~\cite{Salam:07:086} algorithms have been performed~\cite{Abramowicz:10:127}. In \ac{DIS} analyses the jets are clustered in the Breit frame of reference, where the transverse momentum \Pt of jets is a measure of the hardness of the underlying \ac{QCD} process. In analyses of jet production in $\gamma^* p$ interactions, the jets are clustered in the laboratory rest frame. 

\subsection{Jet production in DIS}

The cross section for jet production in the Breit frame in \ac{DIS} can be written as a series expansion in powers of \as, 
\begin{flalign}
\sigma_{jet} = \sum_n \as^{\,n}(\mur) \, &\cdot \, \sum_{i=q, \bar{q}, g} \int \ud \xi \,  f_{i/p}(\xi, \muf^2) \label{eq:sigma_jet} \\
\, &\cdot \, C_{i}^{(n)} \left(\xi, \mur^2, \muf^2 \right) \, \cdot \, (1+\delta_{\mathrm{had}}) \, , \nonumber
\end{flalign}
where $f_{i/p}$ are the \acp{PDF} of the proton, and the coefficient functions $C_{i}^{(n)}$ can be calculated in \ac{pQCD} up to some order of $n$. The variables \muf and \mur are the factorisation and renormalisation scales. The last term in \eq.~\eqref{eq:sigma_jet} represents the hadronisation correction, which is due to non-perturbative effects from the hadronisation process. The two Feynman diagrams in the centre and on the right hand side of \fig.~\ref{fig:feyn} show two of the \ac{LO}, i.e.\ \Oa, contributions to the jet cross section. The \ac{BGF} processes introduce a direct sensitivity of the cross section to the gluon density of the proton and dominate the cross section at small values of \Qsq and transverse jet momentum \Pt. With increasing values of \Qsq and \Pt, the \ac{QCDC} processes become dominant. These are sensitive to \as and the quark distributions and can therefore help to reduce the correlation between the gluon and \as in \ac{QCD} analyses. The coefficient functions $C_{i}^{(n)}$ are known up to \ac{NLO} for dijet and trijet production, resulting in \Oaa calculations for inclusive jet and dijet cross sections and \Oaaa calculations for trijets. The dominating theoretical uncertainty is typically due to missing higher orders in the perturbative series. It is estimated by arbitrary variations of \mur and \muf by conventional factors of 0.5 and 2, and is typically of the order of 10--15\%. 

\subsection{Jets in photoproduction}

In the case of photoproduction the photon can either interact directly with a parton in the proton or it can act as a source of partons, one of which interacts with a parton in the proton. In the latter case \eq.~\eqref{eq:sigma_jet} has to be modified to account for the resolved photon contribution~\cite{Klasen:96:385}. This is achieved by convoluting the coefficient functions with the photon \acp{PDF} in addition to the convolution with the proton \acp{PDF}. The coefficient functions then become $C_{i,j}^{(n)}\left(\xbj, \xgamma, \mur^2, \muf^2 \right)$, where the indices $i$ and $j$ account for the interacting constituents of the proton and the photon. The variable \xgamma is the longitudinal momentum fraction of the photon taken by the parton participating in the hard interaction. In \ac{LO} dijet production, \xgamma can be expressed as 
\begin{equation}
\xgammaobs = \frac{\Ptone e^{-\eta_1} + \Pttwo e^{-\eta_2}}{2 E_\gamma} \, ,
\label{eq:xgamma}
\end{equation}
where $P_{\mathrm{T,i}}$ and $\eta_{i}$ are the transverse momenta and pseudorapidities of the two hard jets, and $E_\gamma$ is the energy of the exchanged photon. In the \ac{LO} picture \mbox{$\xgammaobs = 1$} corresponds to the point-like component of the scattering process. At \ac{NLO} this identification no longer holds, and a sizeable contribution of the hadronic component can be observed~\cite{Frixione:97:315}. In order to relate the $\gamma^* p $ cross section to the total $ep$ cross section, the photon flux is usually calculated using the Weizs\"acker-Williams approximation.

The transition from photoproduction to \ac{DIS} is interesting for a number of reasons. In jet production in $\gamma^* p$ interactions the only hard scale is the jet \Pt, whereas in \ac{DIS} $Q$ may become as hard or even harder than \Pt. The presence of a second hard scale poses an interesting problem for the choice of the renormalisation scale in \ac{pQCD} calculations in \ac{DIS}, which is of no concern in photoproduction. Furthermore, in \ac{DIS} no underlying event is present, and jet production can be studied in a clean environment without contaminations from non-perturbative contributions. In photoproduction the constituents of the resolved photon which did not undergo a hard interaction may scatter from the spectator partons of the proton, such that multiple-interaction models can be studied, and information about the transition from lepton-hadron to hadron-hadron scattering may be obtained. Lastly, in a scenario that is analogous to the \ac{DIS} case, the gluon density of the photon may be constrained by the study of jet production in $\gamma^* p$ interactions.

\section{Determination of \as from jet cross sections}

The H1 and ZEUS collaborations have adopted different methods of extracting the strong coupling \asmz from jet cross sections. In the following, both methods are briefly described and their advantages and disadvantages are summarised.

\subsection{Determination of \asmz by H1}

The H1 collaboration employs the Hessian method~\cite{Barone:00:243,Botje:00:285} for deriving values of \asmz, where the function 
\begin{equation}
\chi^2 = \bm{V}^{T} \bm{M}^{-1} \bm{V} + \sum_k \epsilon_k^2
\label{eq:chi2_h1}
\end{equation}
is minimised. In this definition the matrix $\bm{M}$ is the full correlation matrix, which takes into account the uncorrelated statistical and systematic uncertainties as well as the statistical correlations between different data sets~\cite{Aaron:10:363}. The vectors $\bm{V}$ describe the difference between the experimental data and the theoretical predictions 
\begin{equation}
V_i = \sigma_i^{\mathrm{exp}} - \sigma_i^{\mathrm{theo}} \left( 1 - \sum_k \Delta_{ik} \epsilon_k \right),
\end{equation}
where $\sigma_i^{\mathrm{exp}}$ and $\sigma_i^{\mathrm{theo}}$ are the experimental and predicted cross sections for a given bin $i$, and $\Delta_{ik}$ is the effect of the correlated systematic uncertainty $k$ on the measured cross section in bin $i$. The Hessian parameters $\epsilon_k$ are left free in the fit and used as penalty terms. With this method the experimental uncertainty is given by the natural choice of $\chi^2 = \chi^2_{\mathrm{min}} + 1$.

The theoretical uncertainties due to missing higher orders in the \ac{NLO} calculations, hadronisation corrections and the \ac{PDF} uncertainties are determined with the offset method. In this procedure the theoretical predictions are recalculated with varied input parameters according to their uncertainty and the fit is repeated. The obtained relative change in \asmz is stated as uncertainty. 

\subsection{Determination of \asmz by ZEUS}

The ZEUS collaboration typically performs \ac{QCD} analyses using \ac{NLO} calculations with \ac{PDF} sets that were obtained for different values of \asmz. The dependence of the theory cross section on \asmz as function of an observable $X$ is parametrised in each measurement bin $i$ by a second-order polynomial in \asmz,
\begin{equation}
\frac{\ud \sigma}{\ud X} \Bigg|_i = c_{1,i} \, \asmz + c_{2,i} \, \alpha_s^2(M_Z) \,. 
\end{equation} 
The parameters $c_{n,i}$ are fitted to the theoretical cross sections obtained for different choices of \asmz, where the range $\asmz = 0.115$ to 0.123 is being used. The cross section measured in bin $i$ is mapped to the parametrisation such that the obtained value of \asmz can be obtained easily. The experimental uncertainty on \asmz is obtained by projecting the uncertainty of the measured cross section value on the parametrisation. A combined value of \asmz from $N$ bins is obtained by a $\chi^2$-fit. The correlated systematic uncertainties are taken into account by re-fitting \asmz for all bins with the systematic shifts applied to the measured cross section values~\cite{Tassi:01}.

The theoretical uncertainties due to missing higher orders are estimated by the band method suggested by Jones et al.~\cite{Jones:03:007}. In this method the theoretical predictions are recalculated using the obtained central value of \asmz with the renormalisation and factorisation scales varied up and down by a factor of two. The obtained differences in the prediction are compared to the nominal theory with varying \asmz. The corresponding smallest and largest values of \asmz are then stored for each bin. In a last step the validity of the resulting values of \asmz is tested for each measurement bin and only the allowed values of \asmz are used to set the theoretical uncertainty on \asmz. The uncertainty on \asmz due to the parametrisation of the proton \acp{PDF} is obtained in the same way.

\subsection{Comparison of both methods}

The same \ac{NLO} calculations are used for the \as determinations performed by both experiments. For the \ac{DIS} case calculations based on the Catani-Seymour dipole subtraction~\cite{Catani:97:291} are employed, as implemented in the program NLOJet++~\cite{Nagy:98:14020, Nagy:01:82001}. In the case of photoproduction, calculations using the phase-space-slicing method \cite{Kramer:84} as implemented in the program by Klasen, Kleinwort and Kramer \cite{Klasen:98:1} are used. 

However, as outlined above, the exact methods used by the two experiments for the determination of \asmz are quite different. The H1-type fit takes the systematic uncertainties into account during the fit through the introduction of penalty terms, which allows a cross-check of the determined uncertainties due to systematic mismeasurements. This is not done in the ZEUS method. On the other hand, in the ZEUS case the correlation between \as and the gluon density is correctly taken into account, which is neglected in the case of H1. In the latter case the assumption on the gluon \ac{PDF} can only be tested after the fit by using \ac{PDF} sets obtained with different values of \asmz. 

The theoretical uncertainties are determined in the ZEUS fit based on theory predictions only, whereas in the offset method employed by H1 the experimental uncertainties enter the fit when estimating them. Together with the band method this results in a smaller theoretical uncertainty by about 30--50\% \cite{Aaron:10:363} in the ZEUS case.

\section{Experimental environment}

\subsection{The HERA collider}

The HERA collider was located in Hamburg, Germany and has been in operation between 1992 and 2007. After 15 years of successful operation the two colliding beam experiments H1 and ZEUS each have collected about 0.5~fb$^{-1}$ of $ep$ scattering data. Between the years 2001--2003 a machine upgrade has been carried out, which brought an increase of the instantaneous luminosity by a factor of 4--5 as well as the possibility of longitudinally polarised beams. This shutdown also has been used to replace and upgrade some of the sub-detectors of the experiments. At the end of the HERA-1 phase as well as during the HERA-2 phase the beam energies were 27.6~GeV for the electron\footnote{The HERA collider was operated with electrons and positrons, but since the charge of the lepton beam is of no importance for the physics discussed here, the term electron will be used generically for electrons and positrons.} beam and 920~GeV for the proton beam, resulting in a centre-of-mass energy of about 318~GeV.

\subsection{The H1 and ZEUS experiments}

The two colliding-beam experiments H1 and ZEUS were multi-purpose particle detectors with an asymmetric instrumentation, which reflected the fact that the hadronic centre-of mass system was boosted in the proton direction. Both were equipped with silicon strip detectors close to the interaction point and large jet drift chambers inside a magnetic field of 1.16 and 1.43~T, for the H1 and ZEUS case, respectively. The magnetic field was provided by superconducting coils in both experiments, with the difference being the placement of them. In the case of H1 the superconducting coil surrounded the calorimeter stacks, while in the ZEUS case it was placed between the tracking detectors and the calorimeters. However, the largest difference between the two detectors were the calorimeters employed. H1 used a high-granularity liquid-argon calorimeter with lead and steel plates as absorbers, whereas ZEUS was equipped with a compensating calorimeter based on depleted uranium as absorber and scintillating fibres for the signal detection. This difference is reflected in the achieved energy resolutions. While H1 achieved a better resolution for electrons with $11\% / \! \sqrt{E}$ compared to $18\% / \! \sqrt{E}$ in the ZEUS case, for hadrons ZEUS achieved a better resolution with $35\% / \! \sqrt{E}$ compared to $55\% / \! \sqrt{E}$ for H1, with all resolutions obtained from test beam data. In both experiments large area muon chambers surrounded the calorimeters for the rejection of cosmic ray background, muon identification and the measurement of energy leakage from the calorimeters.

\section{Event reconstruction and jet measurement}

In \ac{DIS} the kinematics can be completely determined by the measurement of the scattered electron. Therefore, the electron identification and reconstruction is crucial for \ac{NC} measurements. The H1 and ZEUS collaborations have achieved excellent precision of the electron energy measurement with experimental uncertainties of the order of 1\%. Also for jet measurements in \ac{DIS} the electron reconstruction is crucial, since the scattered electron is used for the reconstruction of the event kinematics and the determination of the boost to the Breit frame. 

In jet measurements the dominating experimental uncertainty is usually the jet energy scale uncertainty due to the steeply falling cross section as function of jet \Pt. In the following the jet reconstruction for the two experiments is reviewed. 

\subsection{Jet reconstruction at ZEUS}

The compensating calorimeter of the ZEUS experiment with its high resolution for hadronic showers was ideally suited for jet measurements. However, due to the material from the beam pipe, tracking system and superconducting coil in front of the calorimeter on average about 20\% of the jet's transverse energy was lost and needed to be corrected for \cite{Wing:02:767}. This has been achieved by deriving dead-material corrections from \ac{MC} simulations in photoproduction samples using calorimetric and tracking information~\cite{Chekanov:02:9}. Residual corrections were obtained by balancing the scattered electron with the hadronic final state in \ac{NC} events, individually for data and simulated event samples \cite{Chekanov:02:615}. With these corrections applied, a jet energy scale uncertainty of 1\% for jets with $\Pt>10$ GeV has been achieved for ZEUS already in the year 2002. The recent jet measurements by the ZEUS collaboration described in this document use only calorimetric information for the jet reconstruction.  

\subsection{Jet reconstruction at H1}

Due to the non-compensating nature of the H1 calorimeter the energy calibration is more involved and only recently a jet energy scale uncertainty of 1\% has been achieved. The energy calibration procedure is performed in several steps. 

In the first step a software weighting is applied, which corrects for the large fluctuations of the invisible energy fraction in hadronic showers. This $\pi^0$-weighting improves the energy reconstruction and resolution by weighting energy deposits depending on the energy density in single calorimeter cells~\cite{Andrieu:93:499}. However, the calibration coefficients for this method were obtained from \ac{MC} simulations where the absolute energy scale could not be sufficiently well simulated. This resulted in differences between data and \ac{MC} simulations of about 4\%. 

\begin{figure}
\begin{center}
\hspace{-0.2cm}
\includegraphics[width=4cm]{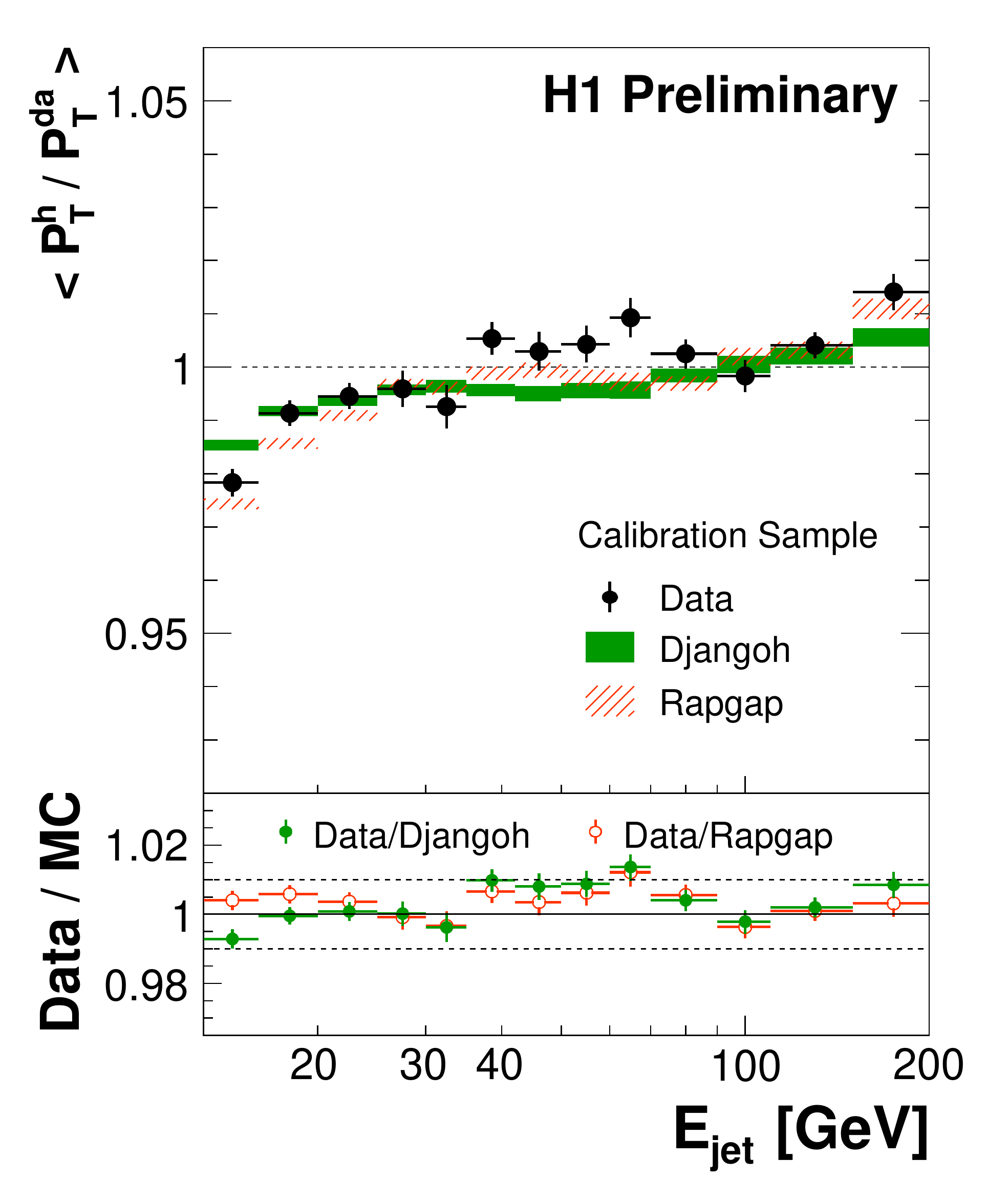} \hspace{-0.2cm} \put(-80,120){\footnotesize (a)}
\includegraphics[width=4cm]{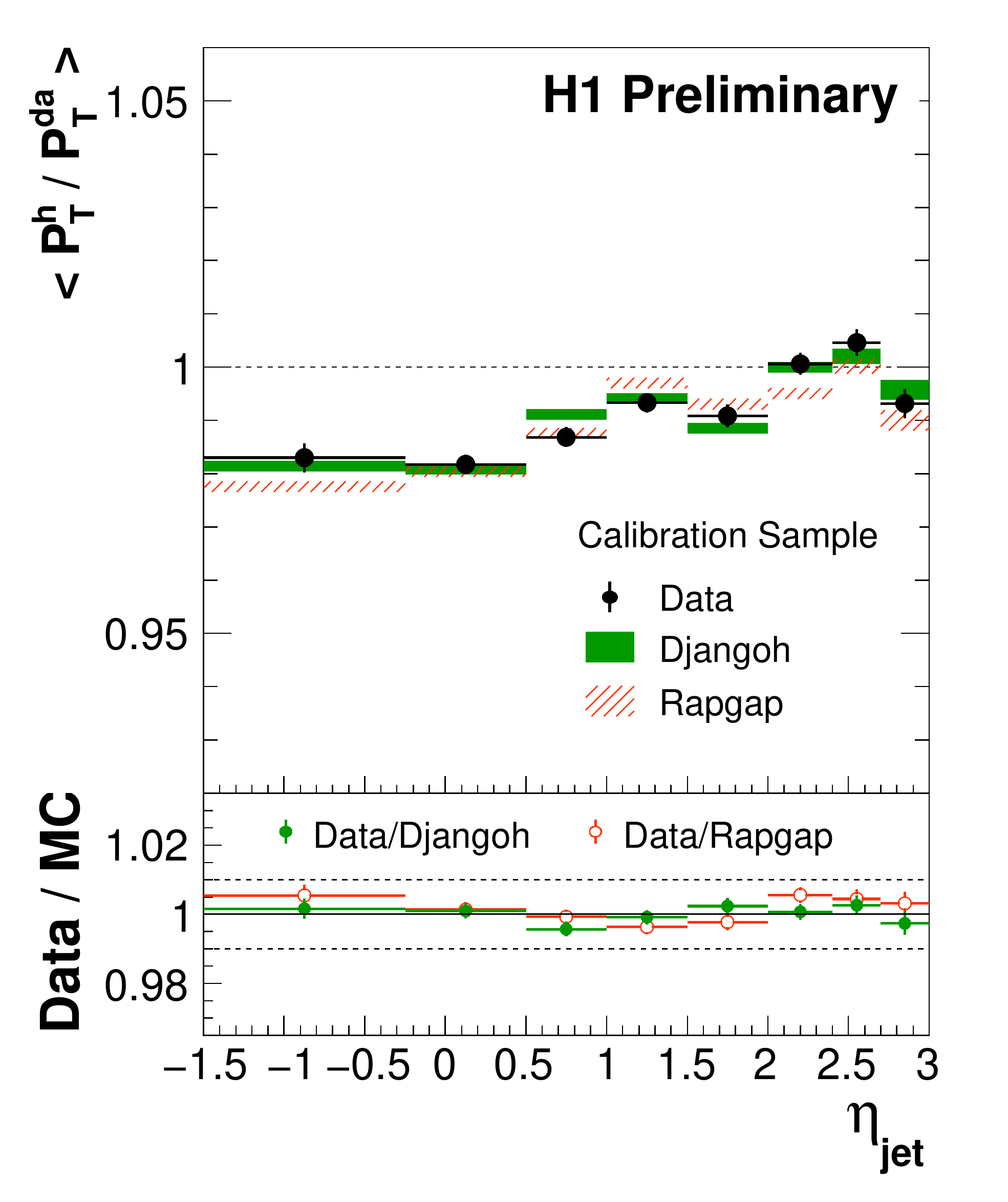} \hspace{-0.2cm} \put(-80,120){\footnotesize (b)}
\vspace{-0.5cm}
\end{center}
\caption{\label{fig:h1_calibration} Ratio of the transverse momentum of the calibrated hadronic final state \Pth to the reference measurement \Ptda as function of the jet energy (a) and the pseudorapidity (b) as obtained by the H1 collaboration. The ratio of the data to \ac{MC} comparison is shown at the bottom of each plot.}
\end{figure}
In order to improve the energy reconstruction an energy-flow algorithm was developed~\cite{Peez:03, Portheault:05}. This algorithm avoids double-counting of energy by comparing the momentum resolution of a reconstructed track with the expected resolution of a matching energy deposit in the calorimeter. If the resolution of the reconstructed track is better, an amount of energy compatible with the track measurement is subtracted from the calorimetric measurement. Otherwise the track is discarded and the calorimetric measurement is kept. 

Recently this algorithm has been improved by separating calorimetric energy deposits originating from electromagnetic and hadronic showers with the help of neural networks~\cite{Kogler:10}. Based on the obtained electromagnetic probability a calibration method has been developed which can be applied independently to data and simulated events. Calibration constants for individual calorimeter clusters have been obtained as a function of their energy and angular region~\cite{Kogler:10}. A comparison of the obtained transverse momentum of the calibrated hadronic final state with a reference measurement is shown in \fig.~\ref{fig:h1_calibration} for data and simulated events~\cite{H1prelim-11--032}. An absolute energy reconstruction of 2\% and an agreement between data and \ac{MC} simulation of 1\% is achieved over the full range of accessible jet energies and pseudorapidities.

\section{Jet production at low \Qsq}

\begin{figure}
\begin{center}
\includegraphics[width=8cm, clip=true, trim=0cm 0cm 0cm 7.6cm]{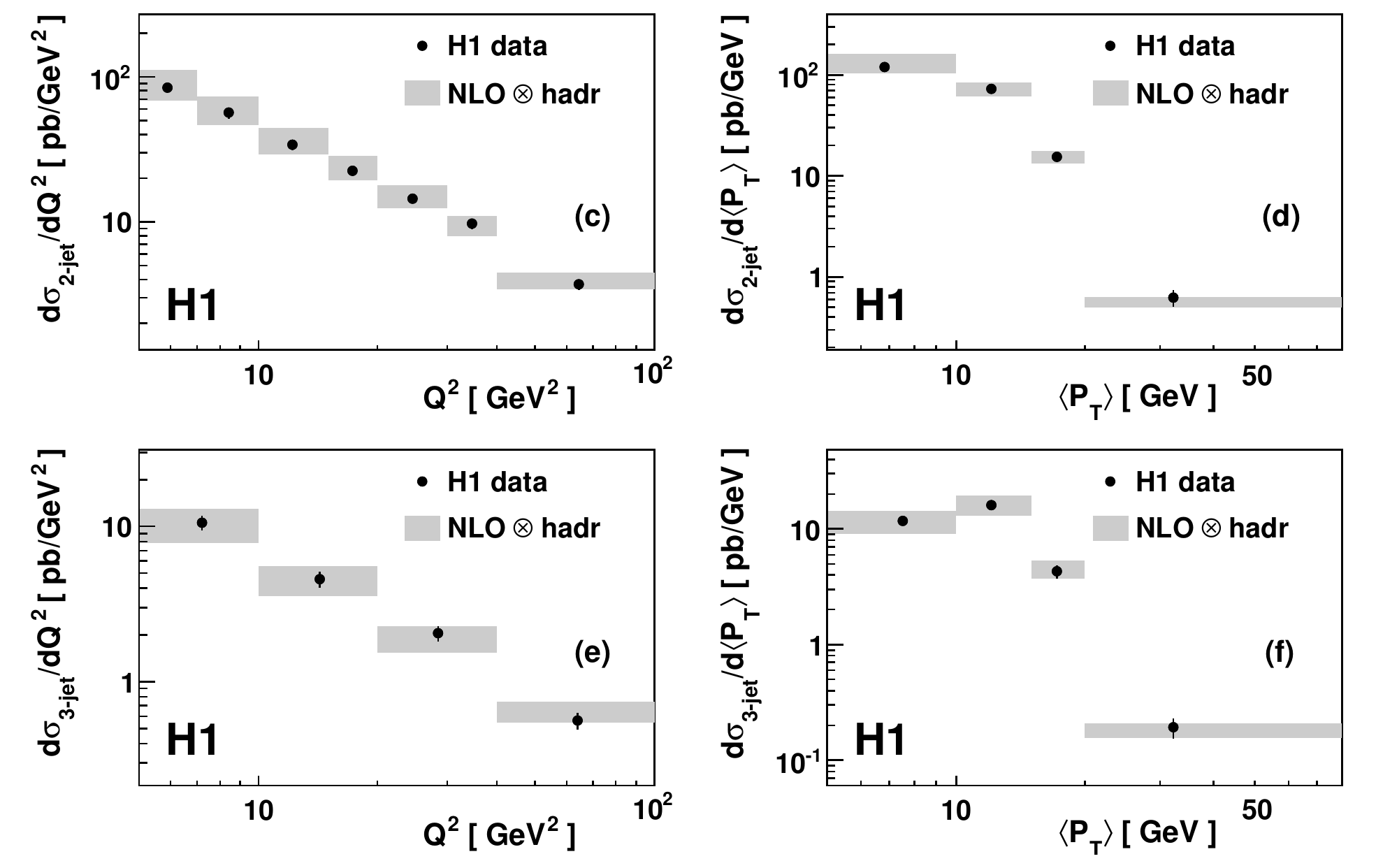} 
\put(  -22, 35){\colorbox{white}{$\phantom{a}$}} \put(  -22, 35){\footnotesize (d)}
\put(-137, 35){\colorbox{white}{$\phantom{a}$}} \put(-137, 35){ \footnotesize (c)}
\put(  -22, 105){\colorbox{white}{$\phantom{a}$}} \put(  -22, 105){ \footnotesize (b)}
\put(-137, 105){\colorbox{white}{$\phantom{a}$}} \put(-137, 105){ \footnotesize (a)}
\end{center}
\vspace{-0.5cm}
\caption{\label{fig:jets_lowq2} The measured cross section for inclusive dijet (a), (b) and trijet production (c), (d) at low \Qsq as function of \Qsq and the average transverse momentum \MeanPt by the H1 collaboration. The measurements are compared to \ac{NLO} calculations corrected for hadronisation effects.}
\end{figure}
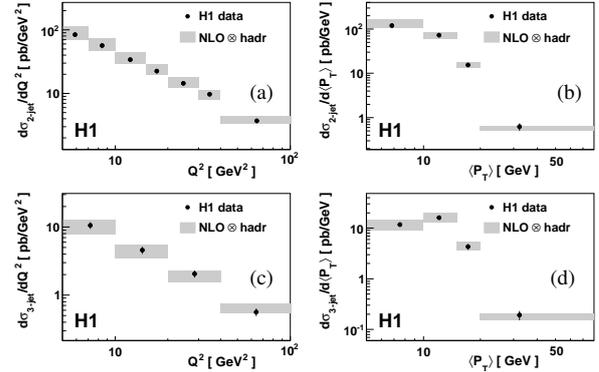
In a recent analysis by the H1 collaboration jet production in the Breit frame is measured at low virtuality of the exchanged boson, $5<Q^2<100$ GeV$^2$~\cite{Aaron:10:1}. In this analysis data from the HERA-1 running phase is used, however, at low \Qsq the cross section is large such that even at high jet \Pt the statistical uncertainties are small. The high statistics allowed for a double-differential measurement of inclusive jet, dijet and trijet production as function of \Qsq and jet \Pt (the average transverse momentum \MeanPt in case of the dijet and trijet measurements). An invariant mass cut of $\Mjj>18$~GeV for the two jets with highest \Pt in the event is required for the dijet and trijet measurements in order to obtain reliable predictions from \ac{NLO} calculations. The measured dijet and trijet cross sections as function \Qsq and \MeanPt in the Breit frame are compared to \ac{NLO} calculations in Figure \ref{fig:jets_lowq2}. The turnover of the trijet distribution at \MeanPt around 12~GeV is due to the cut on the invariant mass and is well described by the \ac{NLO} calculations. The \ac{NLO} calculations are obtained with the choice $\mur = \sqrt{(\MeanPt^2 + \Qsq)/2}$. As an alternative, the choice $\mur=\MeanPt$ has been tested but it is disfavoured by the data in regions where $Q$ is larger than \MeanPt. The theoretical uncertainties, mostly due to the variation of \mur, are large in the phase space of this measurement. They are of the order of 30\% at low \MeanPt and about 10\% at high \MeanPt, whereas the total experimental uncertainties are about 6--10\%. 

As a first step, values of \asmz are extracted for each of the 62 measured cross section points for inclusive jet, dijet and trijet production individually. Good agreement between the obtained values is observed. In a second step a combined fit to all 62 data points is performed using \eq.~\eqref{eq:chi2_h1}, resulting in a value of 
\begin{flalign*}
\asmz = 0.1160 &\pm 0.0014 \, \mathrm{(exp.)} \\
&\pm 0.0016 \, \mathrm{(pdf)} \; ^{+0.0093}_{-0.0077} \, \mathrm{(theo.)} \, ,
\end{flalign*}
with a good fit quality of $\chi^2 / \mathrm{ndf} = 49.8 / 61$. The large theoretical uncertainty is a consequence of the large uncertainty due to missing higher orders in the \ac{NLO} calculation. Smaller theoretical uncertainties can be obtained by studying jet production at high \Qsq.

\section{Jet production at high \Qsq}

The H1 and ZEUS collaborations have measured jet production for \Qsq values larger than 150 and 125 GeV$^{2}$, respectively. Similarly as in the low \Qsq case, inclusive jet, dijet and trijet production cross sections have been measured with high precision, made possible by the larger data samples available from the HERA-2 running phase. 

\subsection{Inclusive jet production}

Preliminary results on inclusive jet production at high \Qsq using HERA-2 data have been released by the H1 and ZEUS collaborations \cite{H1prelim-11--032, ZEUS-prel-10--002}. While the range covered in \Qsq is similar in both analyses, the additional requirements on the phase space in which the measurements are performed are quite different. While the H1 collaboration restricts the inelasticity to $0.2<y<0.7$ and the pseudorapidity range of the jets as measured in the laboratory rest frame to $-1.0 < \etalab < 2.5$, in the ZEUS analysis the measurement is performed for $|\cos\gamma_h|<0.65$ and the jet pseudorapidity requirement in the Breit frame $-2.0< \eta_{\mathrm{Breit}} < 1.5$. The variable $\gamma_h$ is the scattering angle of the hadronic final state and corresponds to the polar angle of the scattered quark in the \ac{LO} picture of \ac{DIS}. These requirements are imposed in order to have an experimentally well controlled boost to the Breit frame and jets well within the geometrical detector acceptance. Both collaborations measured the cross section for inclusive jet production double-differentially as function of \Qsq and jet \Pt. 

\begin{figure}
\begin{center}
\vspace{-0.2cm}
\includegraphics[width=8.5cm]{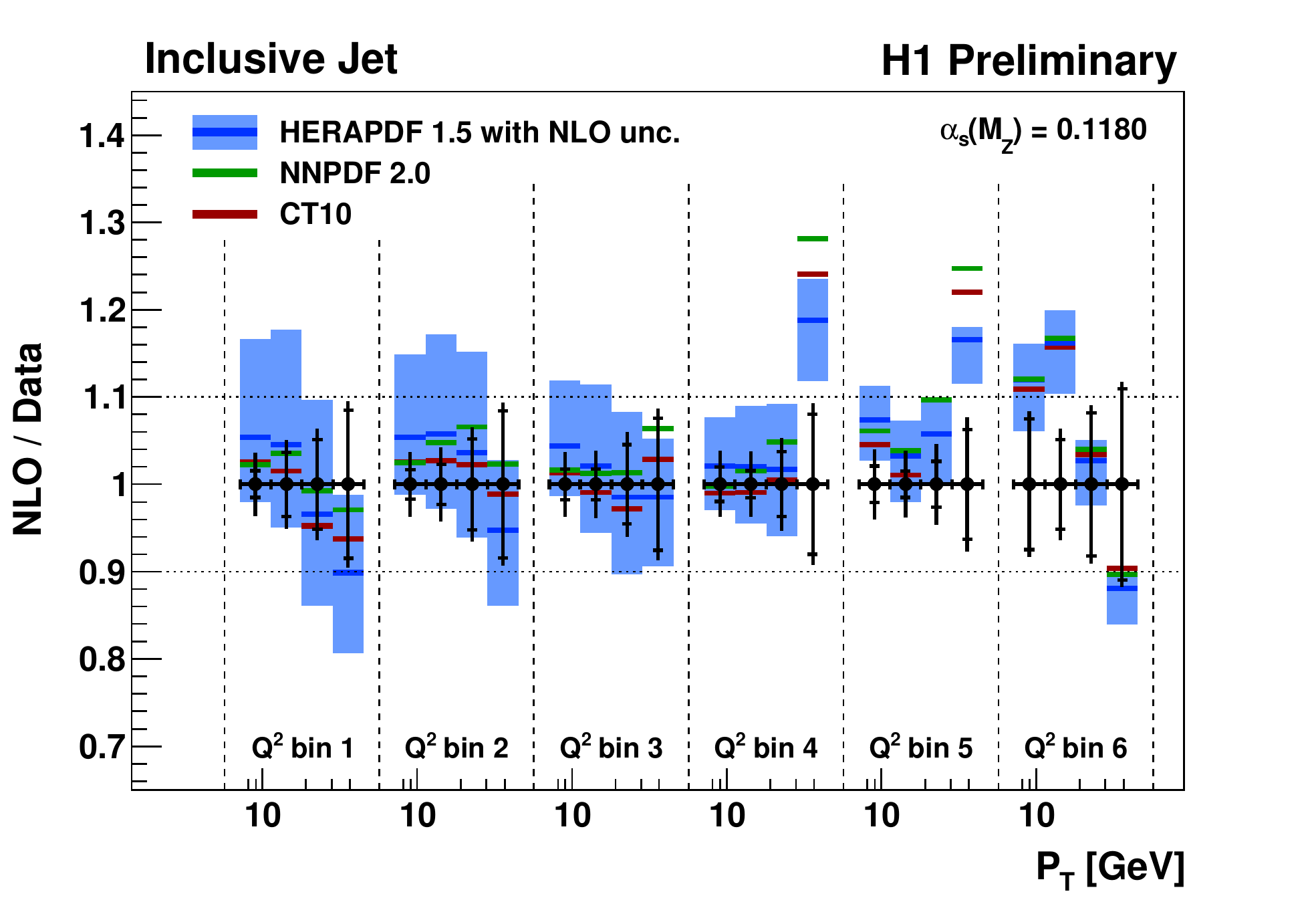}
\vspace{-1cm}
\end{center}
\caption{\label{fig:incjet_highq2} The ratio of \ac{NLO} calculations using different proton \acp{PDF} to the measured cross sections of inclusive jet production at high \Qsq from H1. The theoretical uncertainty due to missing higher orders, estimated by a variation of \mur and \muf, is shown for the calculation using HERAPDF1.5 only.}
\end{figure}
The ratios of the calculated cross sections at \ac{NLO} \ac{QCD}, obtained with different proton \acp{PDF}, to the measured ones by H1 are shown in figure \ref{fig:incjet_highq2}. The \ac{NLO} calculations give a good description of the data. The largest spread of the predictions obtained with different \ac{PDF} sets is observed for $\Pt>30$~GeV for the \Qsq bins 4 and 5, corresponding to $400  <  \Qsq<5000$~GeV$^2$. In this region inclusive jet production is dominated by the valence quark distributions at $0.1<x_p<0.7$, where $x_p$ is the longitudinal momentum fraction of the proton taken by the valence quark, as used by the \ac{NLO} calculations. Since at small values of \Qsq and \Pt the cross section is dominated by gluon-induced processes, the data have the potential to constrain the valence quark at high $x_p$ and provide direct sensitivity to the gluon distribution and \as.

The extraction of \asmz by H1 is performed for all 24 measured bins, and the resulting value is
\begin{flalign*}
\asmz = 0.1190 &\pm 0.0021\,\mathrm{(exp.)} \\
&\pm 0.0020\,\mathrm{(pdf)} \; ^{+0.0050}_{-0.0056}\,\mathrm{(theo.)} \, ,
\end{flalign*}
where the theoretical uncertainty is smaller by nearly a factor of two with respect to the low \Qsq measurement. Further increasing the \Qsq cut for the determination of \asmz improves the theory uncertainty with the drawback of a larger experimental uncertainty. In order to minimise the total uncertainty on \as, the ZEUS collaboration performed a determination of \asmz for $\Qsq>500$~GeV$^2$. The fit is made using four data points of the single-differential $\ud \sigma_{\mathrm{jet}} / \ud \Qsq$ measurement. The resulting value of \asmz is
\begin{flalign*}
\asmz = 0.1208 \; &^{+0.0037}_{-0.0032} \, \mathrm{(exp.)} \\ 
&\!\! \pm 0.0008\,\mathrm{(pdf)} \pm 0.0022 \,\mathrm{(theo.)} \, ,
\end{flalign*}
where the small uncertainty due to the parametrisation of the proton \acp{PDF} is obtained using the \mbox{ZEUS-S}~\cite{Chekanov:03:012007} \ac{PDF} set. The theoretical uncertainty, mostly due to missing higher orders but also including uncertainties of the hadronisation corrections, is only half as large as the corresponding value obtained by H1. 

\subsection{Dijet production}

Although the cross section of inclusive dijet production in the Breit frame is calculated using the same diagrams as inclusive jet production, there are important differences. While in inclusive jet production all jets in a given pseudorapidity region with \Pt above a certain threshold  contribute, an event contributes to the dijet cross section only if at least two jets above a minimum \Pt are found within the given pseudorapidity range. 
This has important consequences. In the inclusive jet case events contribute to the cross section where only one jet lies in the central region and is balanced in transverse momentum by one or more jets with large pseudorapidity. These configurations have larger higher-order corrections, which is reflected in a larger theoretical uncertainty due to the variation of \mur for inclusive jets. 
In the dijet case either an invariant mass or an asymmetric \Pt cut for the two leading jets\footnote{The jet with the highest \Pt in the event is referred to as the leading jet.} is required if the cross section obtained from a fixed-order calculation should be reliable. In the case of a symmetric \Pt cut without a restriction on the invariant mass \Mjj, there are regions in phase space sensitive to soft gluon emission and all-order resummations would be needed~\cite{Frixione:97:315}. 
With either an asymmetric \Pt cut or a requirement on \Mjj fixed-order calculations have been shown to describe dijet data sufficiently. However, a smaller uncertainty due to missing higher orders can be obtained with a cut on the invariant mass. Therefore, the most recent dijet measurements from H1 and ZEUS define the dijet phase space through the requirements $\Pt>5$~GeV and $\Mjj>16$~GeV in the case of the H1 analysis~\cite{H1prelim-11--032} and $\Pt>8$~GeV and $\Mjj>20$~GeV in the case of the ZEUS analysis~\cite{Abramowicz:10:965}. 

\begin{figure}[t]
\hspace{-0.5cm} 
\includegraphics[width=8cm]{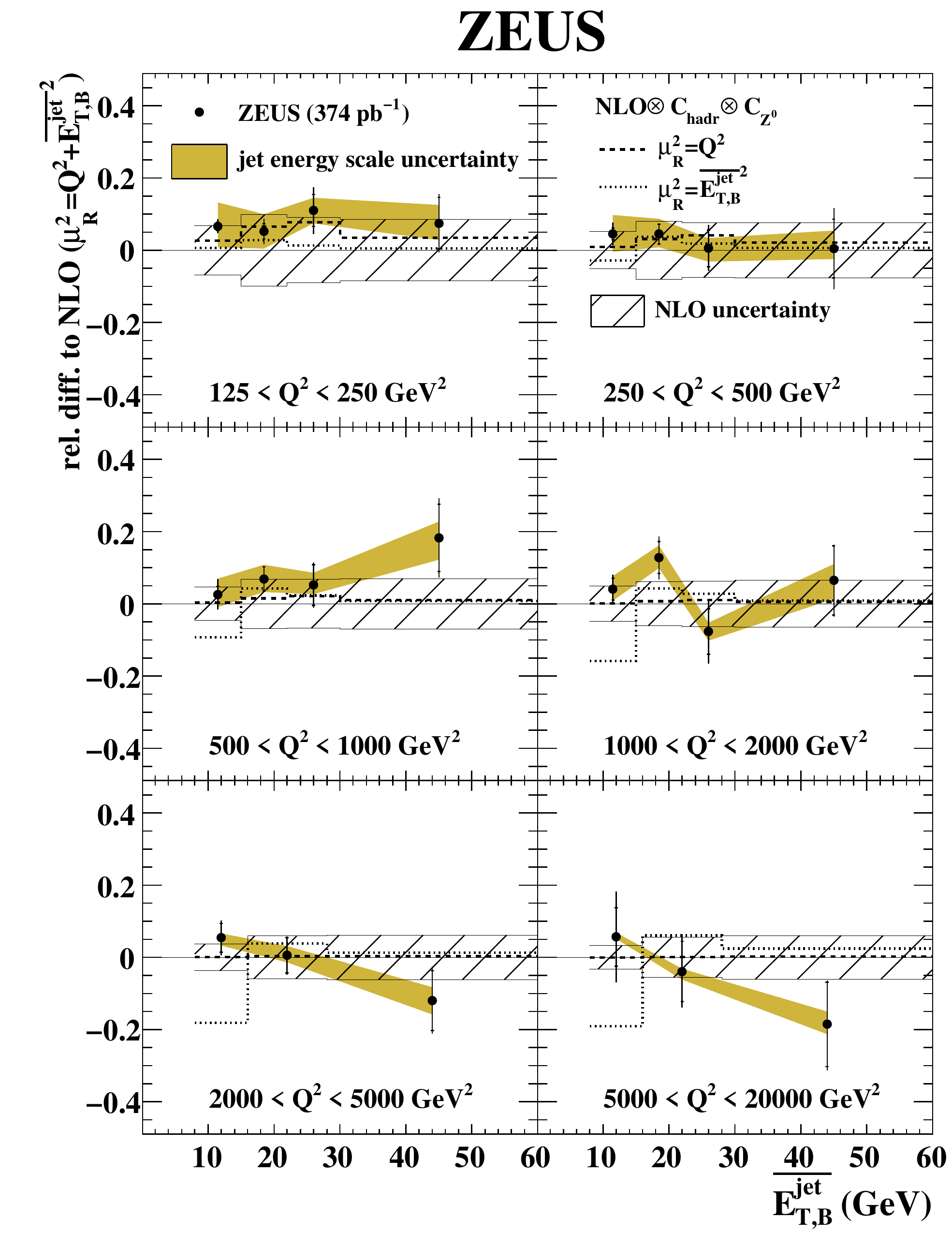}
\vspace{-0.2cm}
\caption{\label{fig:dijet_highq2} The relative differences between the measured double differential dijet cross sections by ZEUS and \ac{NLO} calculations with different choices of \mur. The theoretical uncertainty, shown as dashed area, is only shown for the central prediction with 
$\mur = \sqrt{\Qsq + \MeanPt^2}$. }
\end{figure}
The relative differences between dijet data measured double differentially by the ZEUS collaboration and \ac{NLO} calculations with different choices of the scale \mur are shown in figure \fig.~\ref{fig:dijet_highq2}. A scale choice of \mbox{$\mur = \sqrt{\Qsq + \MeanPt^2}$}, identical to the choice by H1 up to a factor of $1/ \! \sqrt{2}$, leads to a good description of the data over the full region of \Qsq and \MeanPt. The choice of $\mur=\MeanPt$ fails to describe the data for large values of \Qsq when at the same time \MeanPt is small. 

As mentioned above, the theoretical uncertainties due to missing higher orders are smaller in the dijet analysis than in the inclusive jet case. The H1 collaboration determined \asmz from a dijet measurement with a simultaneous fit to 24 bins and obtained a value of 
\begin{flalign*}
\asmz = 0.1146 &\pm 0.0022 \, \mathrm{(exp.)} \\
&\pm 0.0021 \, \mathrm{(pdf)} \; ^{+0.0044}_{-0.0045} \, \mathrm{(theo.)} \, ,
\end{flalign*}
where the theoretical uncertainty is about 20\% smaller than the one obtained for the inclusive jet measurement. 

\subsection{Trijet production}

\begin{figure}[t]
\hspace{-0.3cm} 
\includegraphics[width=8.4cm]{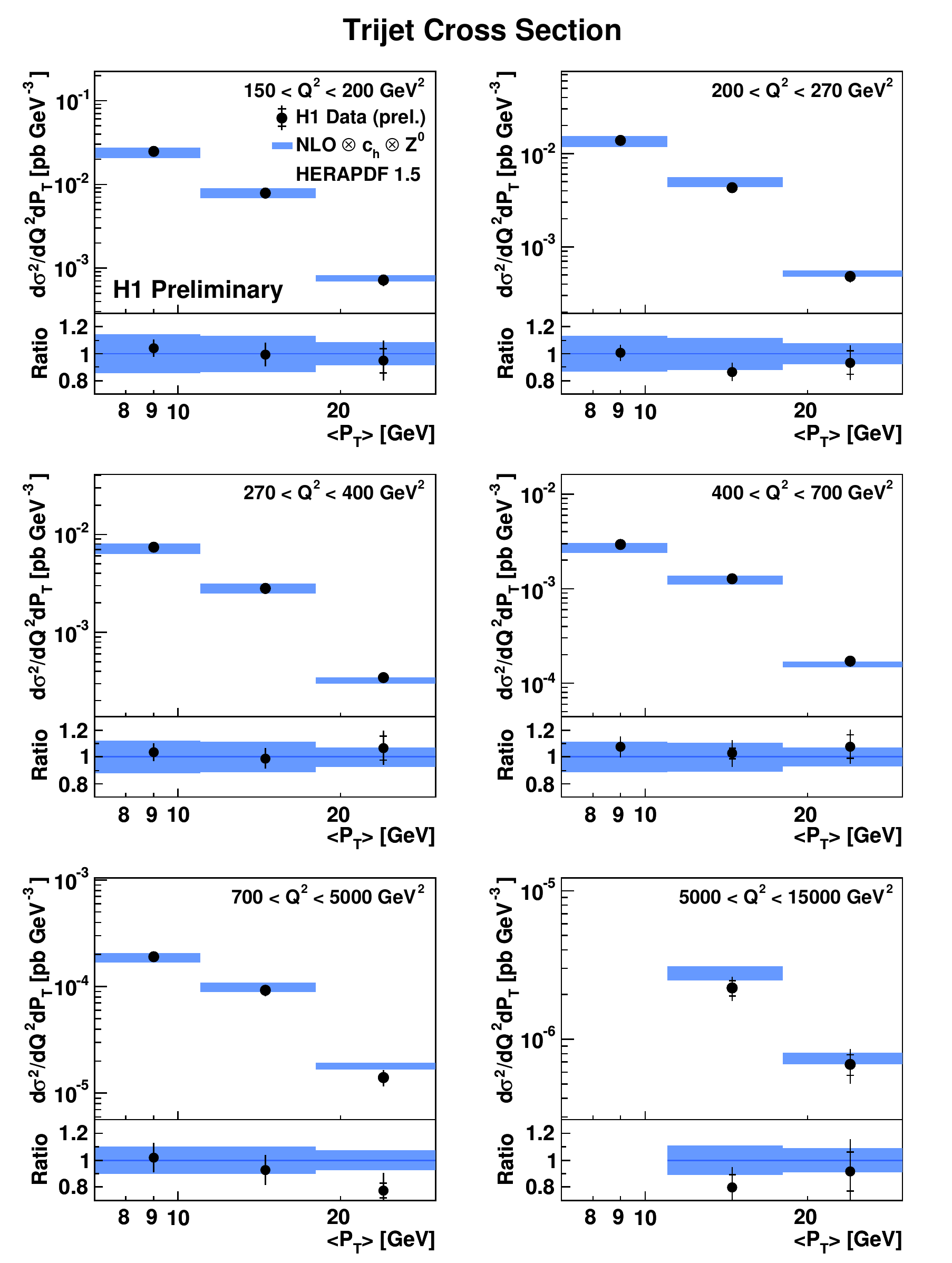}
\vspace{-0.7cm}
\caption{\label{fig:trijet_highq2} Comparison of the trijet cross sections measured by the H1 collaboration with \ac{NLO} calculations using HERAPDF 1.5. The cross sections are shown as function of \MeanPt in different \Qsq bins. The ratio of theoretical predictions to data is shown at the bottom of each plot.}
\end{figure}
The improved reconstruction of the hadronic final state and jet energy scale, together with the high statistics from the HERA-2 running, allowed for the first time a double-differential measurement of trijet production at high \Qsq~\cite{H1prelim-11--032}. In this analysis the same requirement on the invariant mass of the two leading jets as in the dijet analysis is required, $\Mjj>16$~GeV, and all jets are required to have $\Pt>5$~GeV. Because of this, the events used to measure the trijet cross sections are a subsample of the events used for the dijet analysis. This is beneficial for future \ac{QCD} analyses where both measurements are used. 

The measured trijet cross sections have larger experimental uncertainties than the inclusive jet or dijet cross sections, due to larger model uncertainties and a larger effect of the jet energy scale uncertainty. The measurement has total experimental uncertainties between between 7\% at low values of \MeanPt and \Qsq and 15\% at high values of \MeanPt and \Qsq. Also the theoretical uncertainties are larger than in the dijet case by about 50\% and in most bins twice as large as the experimental uncertainties. The \ac{NLO} calculations provide a good description of the trijet data with the choice of $\mur = \sqrt{(\MeanPt^2 + \Qsq)/2}$. A comparison of the measured trijet cross sections with \ac{NLO} calculations using a value of $\asmz = 0.118$ is shown in \fig.~\ref{fig:trijet_highq2}.

The obtained value of \asmz from trijet cross sections has the smallest experimental and \ac{PDF} uncertainty when compared to the values obtained from the inclusive jet and dijet measurements by H1. This is due to the fact that the trijet cross section is already in \ac{LO} proportional to $\alpha_s^2$. The value of \asmz obtained is
\begin{flalign*}
\asmz = 0.1196 &\pm 0.0016 \, \mathrm{(exp.)} \\
&\pm 0.0010 \, \mathrm{(pdf)} \; ^{+0.0055}_{-0.0039} \, \mathrm{(theo.)} \, ,
\end{flalign*}
being in good agreement with the values obtained from the inclusive and dijet analyses. 

\subsection{Normalised jet cross sections}

One way to improve the experimental precision further is to measure ratios of cross sections. For jet production in \ac{NC} \ac{DIS} the ratio of jet cross sections to the inclusive \ac{NC} cross section is an obvious choice. These normalised jet cross sections can be understood as jet rates in \ac{DIS}. The advantage of these observables lies in the cancellation of some experimental uncertainties such as the trigger uncertainty and the uncertainty of the luminosity measurement. Some other uncertainties such as the model dependence of the acceptance correction or the uncertainties related to the electron measurement may cancel partially in the ratio, leading to a further improvement of the experimental precision. This has been impressively demonstrated by a measurement of normalised inclusive jet, dijet and trijet cross sections~\cite{Aaron:10:363} by the H1 collaboration. A simultaneous fit to all 54 data points gives
\begin{flalign*}
\asmz = 0.1168 &\pm 0.0007 \, \mathrm{(exp.)} \\
&\pm 0.0016 \, \mathrm{(pdf)} \; ^{+0.0046}_{-0.0030} \, \mathrm{(theo.)} \, ,
\end{flalign*}
with a good $\chi^2 / \mathrm{ndf}$ of $65 / 53$. The achieved experimental precision of 0.6\% raises the hope of an uncertainty of $\mathcal{O}(1\%)$ once \ac{NNLO} calculations become available.

\section{Jets in photoproduction}

With the study of jet production in $\gamma^*p$ interactions it is possible to gain information on the partonic content of the photon in addition to \as and the structure of the proton. In contrast to \ac{DIS}, there is only one hard scale involved since $\Qsq \approx 0$~GeV$^2$ in photoproduction. The only available choice for the scales \mur and \muf in the \ac{NLO} calculations is therefore the jet \Pt. Although this simplifies the calculations at first sight, the hadronic structure of the photon makes $\gamma^*p$ interactions similar to hadron-hadron interactions which complicates matters. This also means that additional perturbative and non-perturbative effects such as multiple-interactions can play a role in the calculations. However, the advantage of jet production in $\gamma^*p$ compared to hadon-hadron interactions is the absence of multiple-interactions for the point-like contribution. This allows the study of multiple-interaction models in an interesting transition region. 

\begin{figure}[t]
\begin{center}
\includegraphics[width=8.5cm]{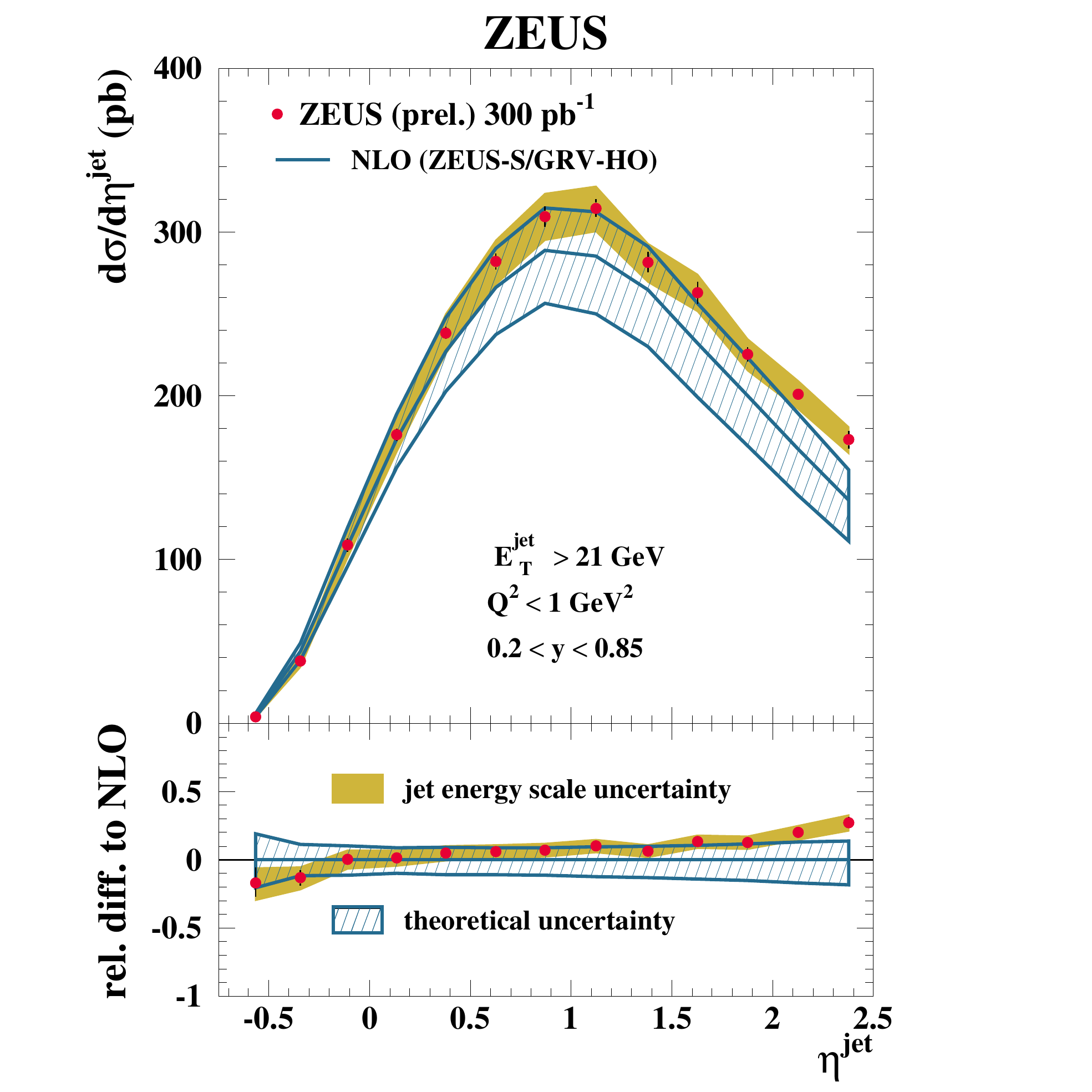}
\end{center}
\vspace{-0.7cm}
\caption{\label{fig:incjet_php_highpt} The inclusive jet cross sections in photoproduction for $\Pt>21$~GeV, measured as function of jet pseudorapidity $\eta^{\mathrm{jet}}$. The \ac{NLO} calculations are corrected for hadronisation effects and shown with the total theoretical uncertainty as hatched band. }
\end{figure}

Recently the ZEUS collaboration measured inclusive jet production in photoproduction~\cite{ZEUS-prel-10--014}, using the full HERA-2 data. In this analysis the selected events are restricted to $\Qsq \le 1$~GeV$^2$, and the $\gamma p$ centre-of-mass energies are in the range of $142 < W_{\gamma p} < 293$~GeV, which corresponds to $0.2<y<0.85$. The selected jets are required to have $\Pt>17$~GeV and have to lie in the pseudorapidity region $-1.0<\etalab<2.5$. The \ac{NLO} calculations are corrected for hadronisation effects by using the ratio of cross sections on the level of stable hadrons to the partonic cross sections using the Pythia and Herwig \ac{MC}. 
\begin{figure}[t]
\begin{center}
\includegraphics[width=8.5cm]{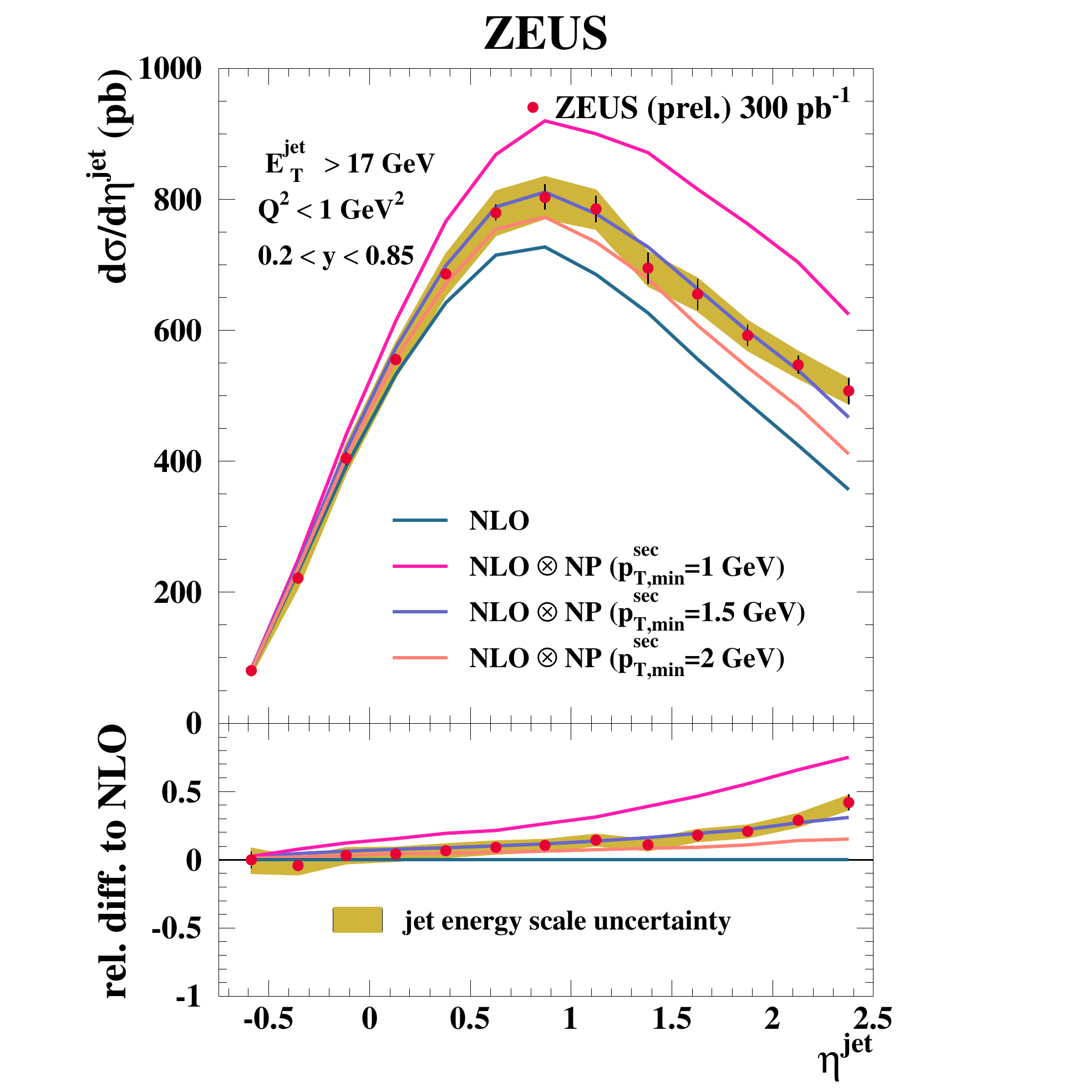}
\end{center}
\vspace{-0.7cm}
\caption{\label{fig:incjet_php} The inclusive jet cross sections in photoproduction as function of jet pseudorapidity $\eta^{\mathrm{jet}}$. The measured cross sections are compared to \ac{NLO} calculations corrected for hadronisation effects and for non-perturbative effects from multiple-interactions~(NP). The latter corrections were obtained with Pythia using different assumptions on the minimum \Pt of the secondary interaction.}
\end{figure}
Disagreements of up to 40\% between the data and the \ac{NLO} predictions can be found at large values of pseudorapidity, $\eta>2$, and small values of jet transverse momenta, $\Pt \lesssim 30$~GeV~\cite{ZEUS-prel-10--014}. This region corresponds to small values of \xgammaobs, see \eq.~\eqref{eq:xgamma}, where the resolved photon is expected to give a large contribution to the cross section and in addition multiple-interactions may occur. In order to study the differences between the data and the predictions in the context of multiple-interactions, an additional correction factor is calculated. It is computed as the ratio of cross sections on the level of stable hadrons using Pythia samples with multiple-interactions and samples without them. The probability of a secondary interaction to occur depends strongly on its allowed value for the minimum transverse momentum $p_{\mathrm{T,min}}^{\mathrm{sec}}$. Three sets of correction factors are obtained for $p_{\mathrm{T,min}}^{\mathrm{sec}} = 1$, 1.5 and 2~GeV. 
The comparison of the modified \ac{NLO} calculations with the data is shown in \fig.~\ref{fig:incjet_php}. While the \ac{NLO} calculations without additional corrections lie significantly below the data, with the disagreement increasing as function of $\eta$, the calculations with an additional correction using $p_{\mathrm{T,min}}^{\mathrm{sec}} = 1.5$ show excellent agreement with the measured cross sections. However, also a different choice of photon \acp{PDF} can reduce the discrepancy between the data and the \ac{NLO} calculations. Increasing the requirement on the minimum jet \Pt reduces the influence of hadronic component of the photon and therefore reduces the contribution of multiple-interactions and the uncertainty due to the photon \acp{PDF}. This behaviour is confirmed in this analysis, where a requirement of $\Pt>21$~GeV for jets results in reasonable agreement between the \ac{NLO} calculations corrected for hadronisation effects only and the data, as shown in \fig.~\ref{fig:incjet_php_highpt}. The total theoretical uncertainty shown as hatched area consists of the uncertainties due to the scale variations, the proton and photon \acp{PDF}, and the hadronisation corrections, added in quadrature. It is larger than the total theoretical uncertainty in the case of the high \Qsq measurements due to the uncertainty from the photon \acp{PDF}, which can be as large as 13\% at high values of $\eta$. 

The value of \asmz is extracted then from the $\ud \sigma / \ud \Pt$ distribution for $21 < \Pt < 71$~GeV, where the upper cut on \Pt is motivated by a relatively large uncertainty due to the proton \acp{PDF} for higher \Pt values. The value obtained is
\begin{flalign*}
\asmz = 0.1206 \; &^{+0.0023}_{-0.0022} \, \mathrm{(exp.)} \\
&\!\! \pm 0.0030 \, \mathrm{(pdf)} \; ^{+0.0042}_{-0.0033} \, \mathrm{(th.)} \, ,
\end{flalign*}
where the uncertainties due to the proton and photon \acp{PDF} have been added in quadrature. The uncertainties of the photon \acp{PDF} have a much larger effect on \asmz than the uncertainties of the proton \acp{PDF}. The relative uncertainty on \asmz related to the photon \acp{PDF} is 2.3\%, which has to be compared with 1\% due to the proton \acp{PDF}. 

\begin{figure}[t]
\begin{center}
\includegraphics[width=8.5cm]{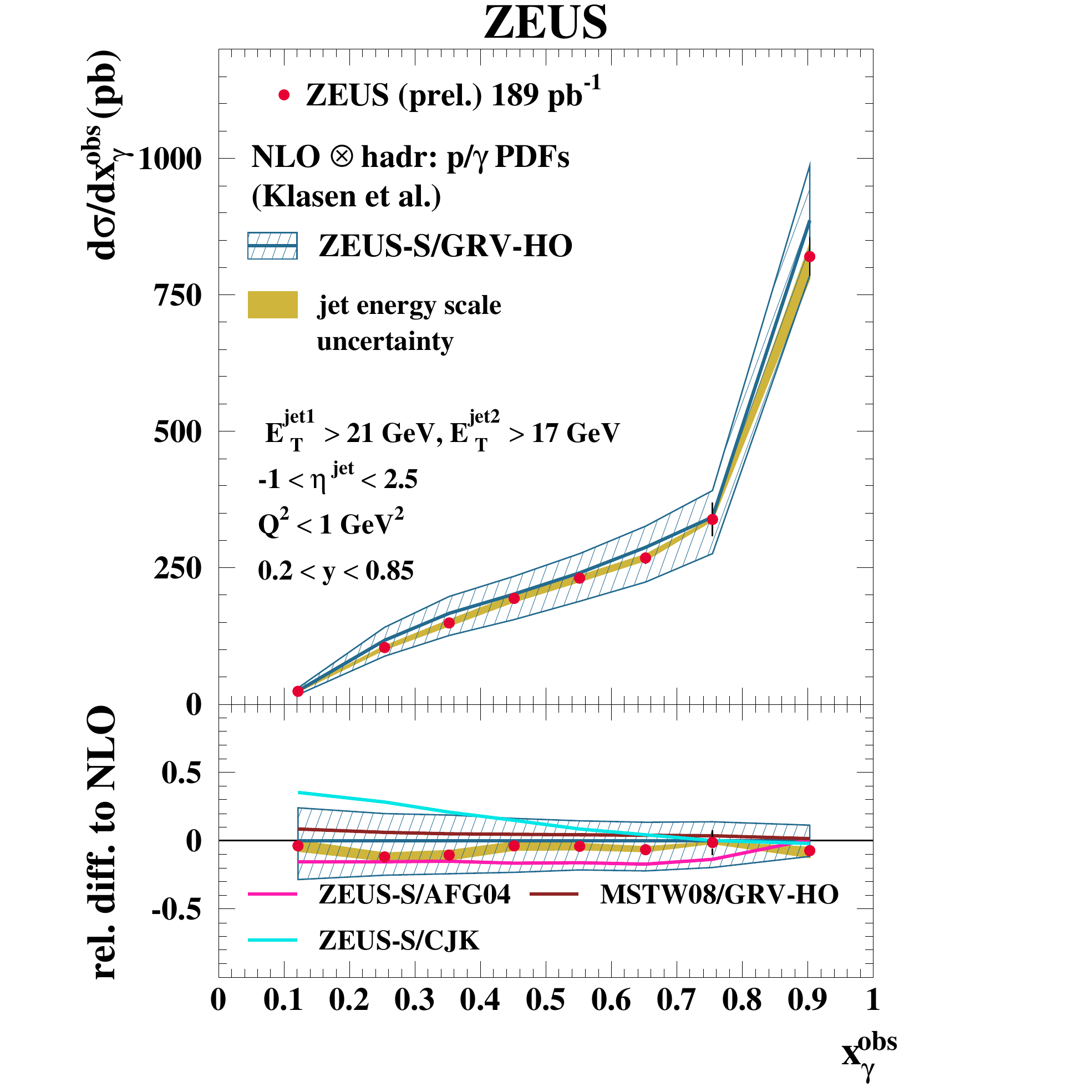}
\end{center}
\vspace{-0.7cm}
\caption{\label{fig:dijet_php} The dijet cross sections in photoproduction as function of \xgammaobs, compared to the \ac{NLO} prediction using three different \ac{PDF} sets for the photon parametrisation. }
\end{figure}
A possibility to further constrain the photon \acp{PDF} and reduce their uncertainties is dijet production in photoproduction, where it is possible to calculate \xgammaobs as defined in \eq.~\eqref{eq:xgamma}. In a new analysis using HERA-2 data, the ZEUS collaboration has measured dijets in $\gamma^* p$ interactions~\cite{ZEUS-prel-11--005} with an asymmetric \Pt requirement of $P_{\mathrm{T,1}} > 21$~GeV and $P_{\mathrm{T,2}} > 17$~GeV. The measured dijet cross sections as function of \xgammaobs are compared to the \ac{NLO} calculation in \fig.~\ref{fig:dijet_php}. The highest jet \Pt in the event is chosen for the scales \mur and \muf in the \ac{NLO} calculations and a good description of the data is observed using the GRV-HO~\cite{Gluck:92:3986, Gluck:92:1973} set of photon \acp{PDF}. At small values of \xgammaobs, as the contribution of the resolved photon becomes more important, a large spread of the predictions using different photon \ac{PDF} sets is observed. Dijet data from HERA may therefore be important for future photon \ac{PDF} determinations and can be used for direct tests of the gluonic content of the photon.

\section{Summary of recent \as values from HERA}

\begin{figure}[t]
\begin{center}
\includegraphics[width=7.5cm]{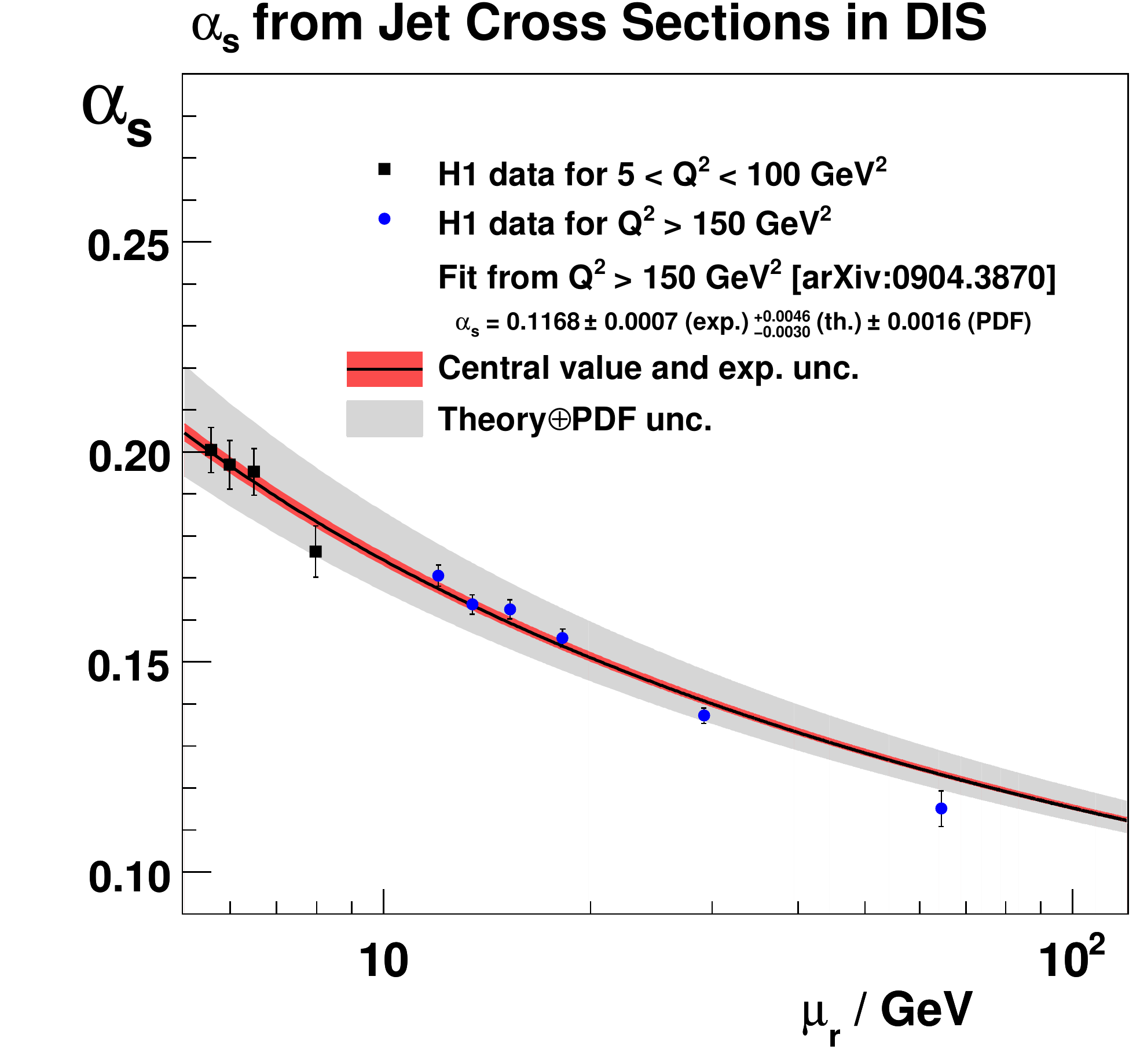}
\end{center}
\vspace{-0.6cm}
\caption{\label{fig:alphas_running_h1} Running of $\as(\mur)$ obtained from multi-jet cross sections in \ac{DIS}~\cite{Aaron:10:1}. The theoretical uncertainty, mostly due to missing higher orders, has been extrapolated using the two-loop solution of the renormalisation group equation from the high \Qsq analysis to the small \Qsq analysis. This explains the relatively small theoretical uncertainty at small values of \mur. }
\end{figure}
A stringent test of \ac{QCD} can be performed by testing the running of the coupling \as at different scales \mur as given by \eq.~\eqref{eq:alphas_running}. Jet data from HERA provide a good possibility for this test, since they cover a large range of scales in a single process. Tests of the running of \as are obtained by directly fitting \asmu, instead of evolving \as to the value of $M_Z$. Results from such a fit are shown in \fig.~\ref{fig:alphas_running_h1}, where \asmu has been fitted to normalised jet data at high \Qsq~\cite{Aaron:10:363} and absolute jet cross sections at low \Qsq~\cite{Aaron:10:1}. For this fit the double differential jet cross sections $\ud^2 \sigma / \ud \Pt \ud \Qsq$ have been used, where the values of \as obtained in the different \Qsq bins have been combined. Very good agreement between the obtained values of \asmu with the predicted running of \as can be observed, where the data cover scales from about 6 to 65~GeV. Interestingly, the \as values obtained in the low \Qsq analysis agree with the extrapolated \asmu from the high \Qsq analysis within the experimental uncertainties. This is quite striking since the theoretical uncertainties in the low \Qsq analysis are twice as large as the ones in the high \Qsq analysis, where only the latter ones are shown as grey band in \fig.~\ref{fig:alphas_running_h1}.

\begin{figure}[t]
\vspace{-1.4cm}
\hspace{-0.35cm}
\includegraphics[width=8.4cm]{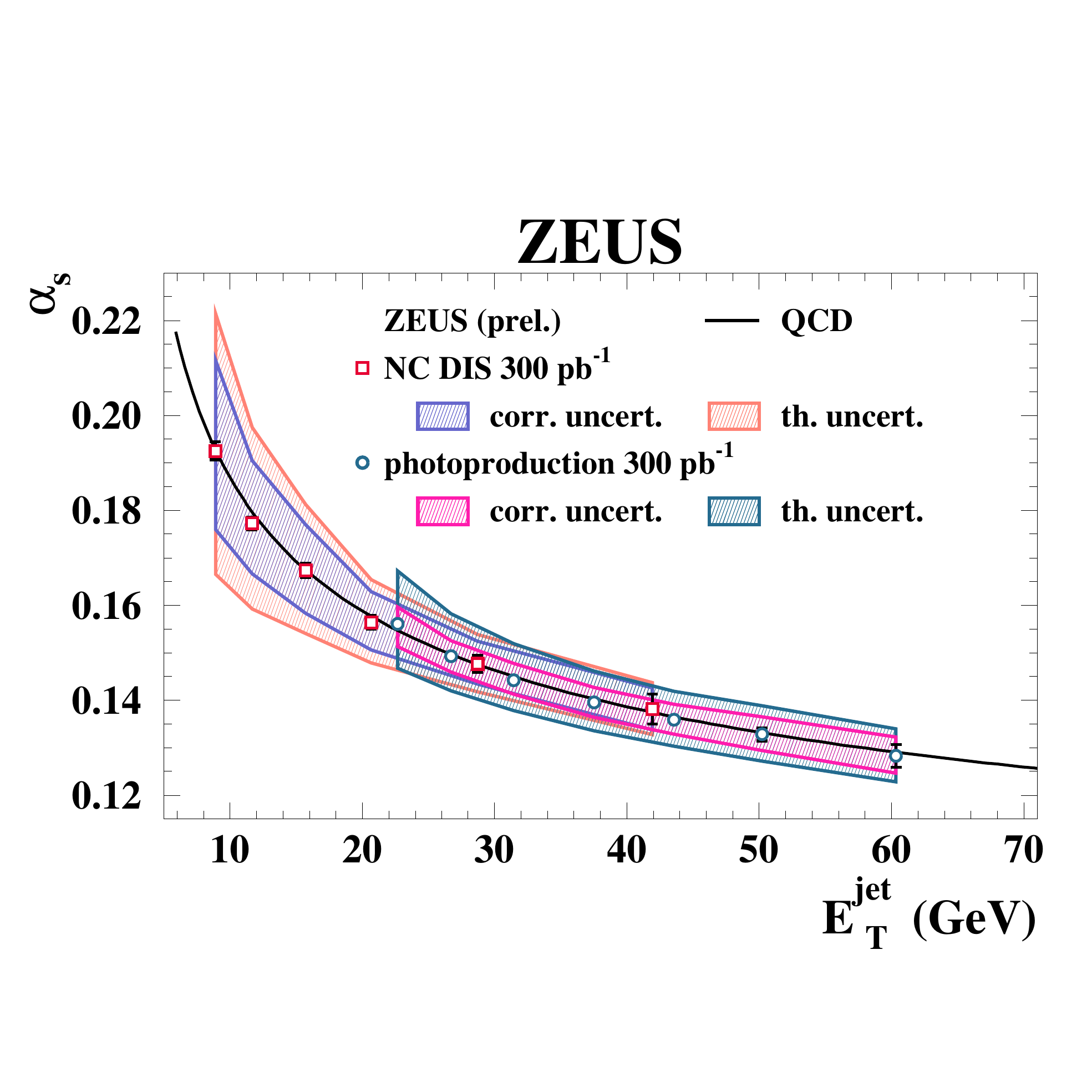}
\vspace{-1.5cm}
\caption{\label{fig:alphas_running_zeus} Running of $\as(\mur)$ obtained from inclusive jet cross sections in photoproduction and \ac{NC} \ac{DIS}~\cite{ZEUS-prel-11--005}. The experimental uncertainties are shown with their uncorrelated and correlated parts. The solid line corresponds to the two-loop solution of the renormalisation group equation with $\asmz=0.1206$.}
\end{figure}
A similar study~\cite{ZEUS-prel-11--005} has been carried out by the ZEUS collaboration, where the running of \asmu is tested with inclusive jet data in photoproduction and high \Qsq \ac{NC} \ac{DIS}. The result is shown in \fig~\ref{fig:alphas_running_zeus}. In this case, the fit has been performed using the measured $\ud \sigma / \ud \Pt$ distributions, and the average jet \Pt measured in each bin has been chosen for the scale \mur. Good agreement between the obtained values of \asmu and the two-loop solution of the renormalisation group equation is observed also in this analysis. The transition between photoproduction and \ac{DIS} is smooth and does not show any peculiarities. 

\begin{figure}[t]
\hspace{-0.4cm}
\includegraphics[width=8.5cm]{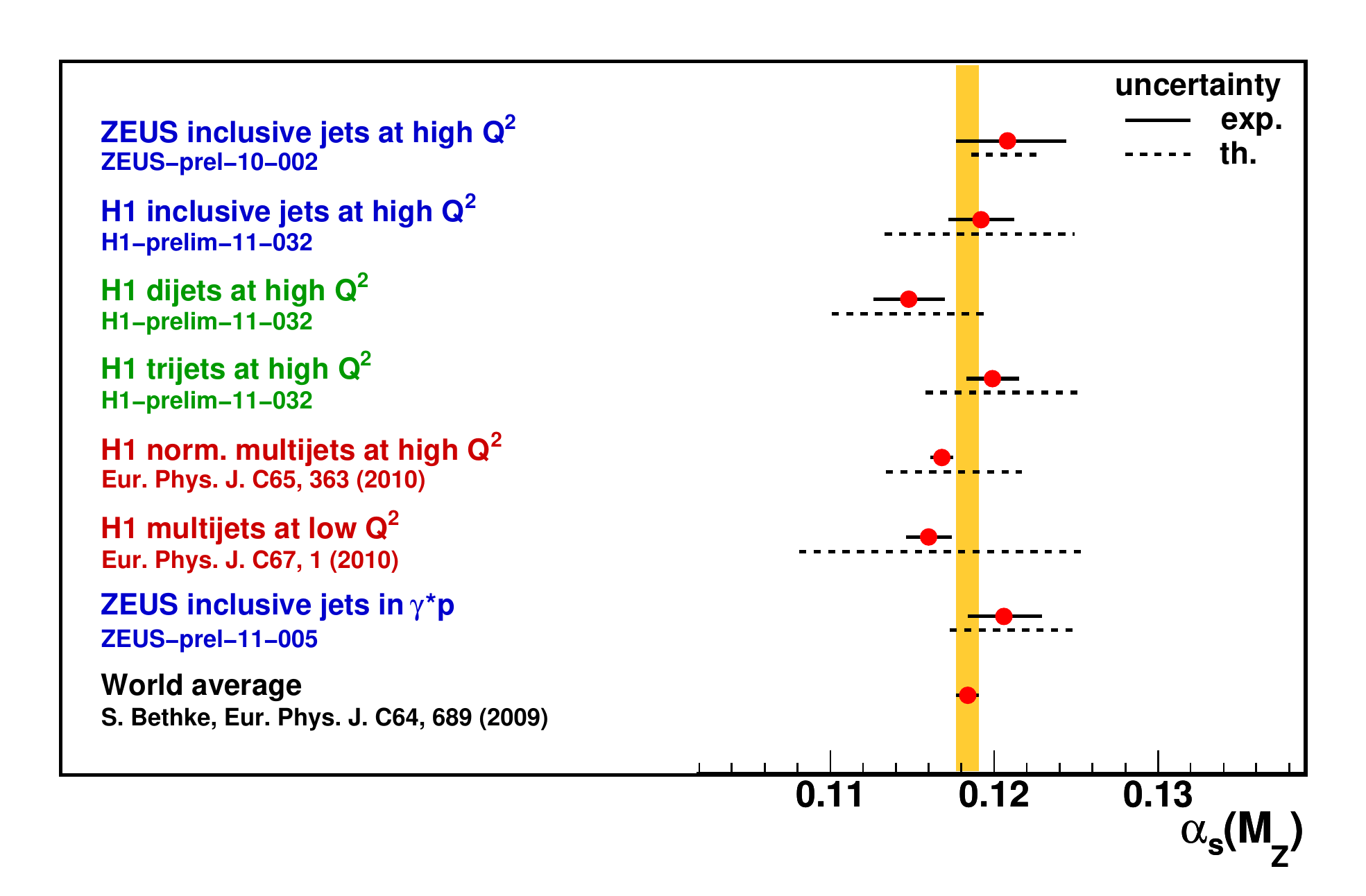}
\vspace{-0.7cm}
\caption{\label{fig:alphas_summary} Comparison of the most recent values of \asmz obtained from multi-jet production cross sections at HERA and the world average~\cite{Bethke:09:689}. }
\end{figure}
A summary of all values of \asmz given in this document is shown in \fig.~\ref{fig:alphas_summary}. The obtained values show good agreement  with the world average~\cite{Bethke:09:689}. The HERA measurements of \asmz from jet data are clearly dominated by the theoretical uncertainties, which are mostly due to missing higher orders. However, the experimental uncertainties are comparable with the precision of the world average and once \ac{NNLO} calculations become available, determinations of \asmz with a precision at the percent level will be feasible using jet data from HERA. It should be noted that the precision of the world average is mostly driven by lattice calculations and the analysis of $\tau$-decays. When comparing the precision of the recent \as determinations from HERA jet data with other collider data, the achieved precisions are competitive. 

\section{Conclusion}

The last few years have brought a large improvement in terms of experimental precision for jet measurements from HERA. This is partly due to the higher statistics of the HERA-2 data, but mostly because of achievements of the experimental collaborations in terms of their reconstruction and calibration algorithms. Both, the H1 and ZEUS collaborations have reached the goal of a 1\% jet energy scale uncertainty. They are now in the process of finalising their physics programmes which include high precision jet measurements, with the first ones published recently.

These measurements include analyses of inclusive jet, dijet and trijet production in different kinematic regimes and provide stringent tests of \ac{pQCD}. Even more importantly, these data will be important for future \ac{QCD} analyses. The potential of these data can already be seen by the precision determinations of \asmz with experimental uncertainties of $\mathcal{O}(1\%)$. However, the theoretical uncertainties, which are mostly due to missing higher orders, are limiting the precision of the extracted values of \asmz to about 3--4\%. The obtained values of \asmz from the different analyses show excellent agreement. This is quite remarkable considering that they have been obtained using different methods and data in very different kinematic regimes from two different experiments. 

This opens the possibility for a determination of \asmz using the latest jet data from H1 and ZEUS simultaneously, which may improve the experimental precision further. The last attempt in this direction has been made in 2007 and showed a reduction of the experimental uncertainty of about 30\% with respect to the individual determinations~\cite{H1prelim-07--132}. 

Another interesting possibility presents itself by a combination of jet cross sections in a similar way as the combination of inclusive \ac{DIS} data~\cite{Aaron:10:109}. There the final cross sections are obtained by a fit to the individual data sets. In this way not only the statistical uncertainties improve, but also the systematic uncertainties are potentially reduced because the two experiments are allowed to cross-calibrate one another. This approach would have the advantage of providing jet data with the ultimate precision, which can be used in future \ac{QCD} analyses with \ac{NNLO} accuracy.



\bibliographystyle{h-elsevier}
\bibliography{kogler}

\begin{thebibliography}{10}

\bibitem{H1prelim-11--034}
H1 and ZEUS Coll.,
\newblock Preliminary result, H1prelim-11-034, ZEUS-prel-11-001, 2011.

\bibitem{Ritbergen:97:379}
T.V. Ritbergen, J. Vermaseren and S. Larin,
\newblock Phys. Lett. B 400 (1997) 379.

\bibitem{Watt:11:69}
G. Watt,
\newblock JHEP 2011 (2011) 69, see also these proceedings.

\bibitem{Aaron:10:109}
H1 and ZEUS Coll., F. Aaron et~al.,
\newblock JHEP 1001 (2010) 109.

\bibitem{HERAPDF}
K. Lipka,
\newblock these proceedings.

\bibitem{Catani:93:187}
S. Catani et~al.,
\newblock Nucl. Phys. B 406 (1993) 187.

\bibitem{Ellis:93:3160}
S. Ellis and D. Soper,
\newblock Phys. Rev. D 48 (1993) 3160.

\bibitem{Huth:90}
J. Huth et~al.,
\newblock Research directions for the decade. Summer Study on HEP, Snowmass,
  Colorado  (1990).

\bibitem{Cacciari:08:063}
M. Cacciari, G. Salam and G. Soyez,
\newblock JHEP 04 (2008) 063.

\bibitem{Salam:07:086}
G. Salam and G. Soyez,
\newblock JHEP 05 (2007) 086.

\bibitem{Abramowicz:10:127}
ZEUS Coll., H. Abramowicz et~al.,
\newblock Phys. Lett. B 691 (2010) 127.

\bibitem{Klasen:96:385}
M. Klasen and G. Kramer,
\newblock Phys. Lett. B 366 (1996) 385.

\bibitem{Frixione:97:315}
S. Frixione and G. Ridolfi,
\newblock Nucl. Phys. B 507 (1997) 315.

\bibitem{Barone:00:243}
V. Barone, C. Pascaud and F. Zomer,
\newblock Eur. Phys. J. C 12 (2000) 243.

\bibitem{Botje:00:285}
M. Botje,
\newblock Eur. Phys. J. C 14 (2000) 285.

\bibitem{Aaron:10:363}
H1 Coll., F.D. Aaron et~al.,
\newblock Eur. Phys. J. C 65 (2010) 363.

\bibitem{Tassi:01}
E. Tassi,
\newblock Measurement of Dijet Production in Neutral Current Deep Inelastic
  Scattering at High $Q^2$ and Determination of $\alpha_s$ at HERA,
\newblock Dissertation, Universit\"at Hamburg, 2001.

\bibitem{Jones:03:007}
R. Jones et~al.,
\newblock JHEP 12 (2003) 007.

\bibitem{Catani:97:291}
S. Catani and M. Seymour,
\newblock Nucl. Phys. B 485 (1997) 291.

\bibitem{Nagy:98:14020}
Z. Nagy and Z. Trocsanyi,
\newblock Phys. Rev. D 59 (1998) 14020.

\bibitem{Nagy:01:82001}
Z. Nagy and Z. Trocsanyi,
\newblock Phys. Rev. Lett. 87 (2001) 82001.

\bibitem{Kramer:84}
G. Kramer,
\newblock Theory of Jets in Electron-Positron Annihilation (Springer Verlag,
  Berlin, 1984).

\bibitem{Klasen:98:1}
M. Klasen, T. Kleinwort and G. Kramer,
\newblock Eur. Phys. J. Direct C 1 (1998) 1.

\bibitem{Wing:02:767}
M. Wing,
\newblock Proc. of the 10th Int. Conf. on Calorimetry in HEP, R. Zhu (ed.)
  (2002) 767.

\bibitem{Chekanov:02:9}
ZEUS Coll., S. Chekanov et~al.,
\newblock Phys. Lett. B 531 (2002) 9.

\bibitem{Chekanov:02:615}
ZEUS Coll., S. Chekanov et~al.,
\newblock Eur. Phys. J. C 23 (2002) 615.

\bibitem{Andrieu:93:499}
Calorimeter Group of H1, B. Andrieu et~al.,
\newblock Nucl. Instr. and Meth. A 336 (1993) 499.

\bibitem{Peez:03}
M. Peez,
\newblock Search for deviations from the standard model in high transverse
  energy processes at the electron proton collider HERA (in French),
\newblock Dissertation, Univ. Claude Bernard, Lyon, DESY-THESIS-2003-023,
  CPPM-T-2003-04, 2003.

\bibitem{Portheault:05}
B. Portheault,
\newblock First measurement of charged and neutral current cross sections with
  the polarized positron beam at HERA II and QCD-electroweak analyses (in
  French),
\newblock Dissertation, Univ. Paris XI Orsay, LAL-05-05, 2005.

\bibitem{Kogler:10}
R. Kogler,
\newblock Measurement of jet production in deep-inelastic $ep$ scattering at
  HERA,
\newblock Dissertation, Universit{\"a}t Hamburg, DESY-THESIS-2011-003,
  MPP-2010-175, 2010.

\bibitem{H1prelim-11--032}
H1 Coll.,
\newblock Preliminary result, H1prelim-11-032, 2011.

\bibitem{Aaron:10:1}
H1 Coll., F.D. Aaron et~al.,
\newblock Eur. Phys. J. C 67 (2010) 1.

\bibitem{ZEUS-prel-10--002}
ZEUS Coll.,
\newblock Preliminary result, ZEUS-prel-10-002, 2010.

\bibitem{Chekanov:03:012007}
ZEUS Coll., S. Chekanov et~al.,
\newblock Phys. Rev. D 67 (2003) 012007.

\bibitem{Abramowicz:10:965}
ZEUS Coll., H. Abramowicz et~al.,
\newblock Eur. Phys. J. C 70 (2010) 965.

\bibitem{ZEUS-prel-10--014}
ZEUS Coll.,
\newblock Preliminary result, ZEUS-prel-10-014, 2010.

\bibitem{ZEUS-prel-11--005}
ZEUS Coll.,
\newblock Preliminary result, ZEUS-prel-11-005, 2011.

\bibitem{Gluck:92:3986}
M. Gl{\"u}ck, E. Reya and A. Vogt,
\newblock Phys. Rev. D 45 (1992) 3986.

\bibitem{Gluck:92:1973}
M. Gl{\"u}ck, E. Reya and A. Vogt,
\newblock Phys. Rev. D 46 (1992) 1973.

\bibitem{Bethke:09:689}
S. Bethke,
\newblock Eur. Phys. J. C 64 (2009) 689, see also these proceedings.

\bibitem{H1prelim-07--132}
H1 and ZEUS Coll.,
\newblock Preliminary result, H1prelim-07-132, ZEUS-prel-07-025, 2007.

\end{thebibliography}

\end{document}